\documentclass{aa}  

\usepackage{graphicx}
\usepackage{txfonts}
\usepackage{hyperref}
\begin{document}

   \title{FIRTEZ-dz:}
   \subtitle{A Forward and Inverse solver of the polarized Radiative Transfer Equation under Zeeman regime in geometrical scale}

   \author{A. Pastor Yabar
          \inst{1}
          \and
          J. M. Borrero\inst{1}
          \and
          B. Ruiz Cobo\inst{2,3}
          }

   \institute{Leibniz Institut f\"ur Sonnenphysik,
              Sch\"oneckstr. 6, 79104 Freiburg, Germany\\
              \email{apy@leibniz-kis.de}
         \and
             Instituto de Astrof\'{\i}sica de Canarias,
             C. V\'{\i}a L\'actea s/n., 38205 La Laguna, Tenerife, Spain
         \and
             Departamento de Astrof\'{\i}sica de Canarias, Universidad de La Laguna,
             Avda Astrof\'{\i}sico S\'anchez s/n., 38206, La Laguna, Tenerife, Spain
             }

   \date{Received September 15, 1996; accepted March 16, 1997}

 
  \abstract
{
We present a numerical code that solves the forward and inverse problem of the polarized radiative transfer equation in geometrical scale under the Zeeman regime. The code is fully parallelized, making it able to easily handle large observational and simulated datasets. We checked the reliability of the forward and inverse modules through different examples. In particular, we show that even when properly inferring various physical parameters (temperature, magnetic field components, and line-of-sight velocity) in optical depth, their reliability in height-scale depends on the accuracy with which the gas-pressure or density are known. The code is made publicly available as a tool to solve the radiative transfer equation and perform the inverse solution treating each pixel independently. An important feature of this code, that will be exploited in the future, is that working in geometrical-scale allows for the direct calculation of spatial derivatives, which are usually required in order to estimate the gas pressure and/or density via the momentum equation in a three-dimensional volume, in particular the three-dimensional Lorenz force.
}


   \maketitle
%

\section{Introduction}
\label{sec:introduction}

In observational sciences, meaningful physical properties of an object are inferred from available  observations. In the context of the stellar atmospheres, and in particular for this work, the solar photosphere, most of this inference is performed by means of different approximations to the solution of the radiative transfer equation (hereinafter referred to as RTE) applied to the spectra and polarized spectra emitted by the object.

The RTE characterizes the interaction of the matter with the light that traverses through it. This interaction leaves its `fingerprints' in many different ways, for instance in the shape of the continuum of the spectra coming from the object, or in the presence or absence and the shape of spectral lines. Properly modeled, these features allow approximate characterization of the studied object. In recent decades, the advent of polarimetric observations has allowed the generalization of the RTE to polarized light \citep{unno1956,rachkovsky1967,landi1972}. This step was of great significance, as the polarization of the light encodes information on the magnetic field configuration as well as the geometry of the system, which allows a deeper insight into the system being studied.

In solar physics, there are different approaches to solving the polarized RTE depending on the magnetic field regime adopted: atomic, Hanle, Zeeman, and/or Paschen \citep[see e.g.,][]{landi2004}. In this work, we focus on the solution of the polarimetric RTE under the Zeeman magnetic regime, which is usually the dominant one at this atmosphere layer.

There are already several methods to address the solution of the RTE, each with its respective accuracy \citep[see, e.g.,][]{deltoroiniesta2016}. There are several  computational codes that allow one to infer the optical depth ($\tau$) dependence of the physical properties in the solar atmosphere; for example, SIR \cite{ruizcobo1992}, SPINOR \cite{frutiger2000}, NICOLE \cite{socasnavarro2015}, SNAPI \citep{milic2018}, and STIC \citep{delacruzrodriguez2019}. These are of particular relevance here as they allow the furthest insight into the properties of the solar atmosphere in a three-dimensional volume. In these cases, the conversion between optical depth $\tau$ and geometrical height $z$ is then performed by considering the relation $d\tau=-\rho\kappa_{c}dz$, where $\rho$ is the density and $\kappa_{c}$ the continuum opacity, with the latter being a nonlinear function of the temperature $T$ and gas pressure $P_{g}$. Since the proper calculation of the gas pressure (and density) involves the whole three-dimensional volume, for which there is no account in the aforementioned codes, the gas pressure and density inferred by them are unrealistic, and hence also the conversion between $\tau$ and $z$. Deviations from hydrostatic equilibrium involve, among others, time-variation of the velocity, advection and viscous terms and, the most relevant at this atmospheric layer, the Lorentz force. The latter is responsible for the Wilson depression, a depression of the height at which photospheric spectral lines are formed with respect to the quiet-Sun because of the presence of magnetic field. Because the geometrical scale $z$ is then inaccurate, so is the calculation of the spatial derivatives, such as those involved in the determination of $\nabla\cdot\mathbf{B}=0$ and the electric currents $\mathbf{j}=\frac{1}{4\,\pi}\nabla\times\mathbf{B}$. Consequently, it is not possible to establish whether the inferred magnetic field verifies Maxwell's equations.

Up to now there have only been a handful of attempts to address these problems and limitations. One of them was presented by \cite{puschmann2010}, who used the SIR code to infer the various physical parameters in a $\tau$-scale before transforming these parameters  into a self-consistent geometrical-height scale $z$ in a three-dimensional volume. To that end, these latter authors used a more detailed gas pressure ($P_{g}$) and density ($\rho$) than the one given by the SIR code as they were calculated by minimizing the static equation of momentum in ideal magnetohydrodynamics (MHD) including the Lorentz force, together with the $\nabla\cdot\mathbf{B}=0$ condition. In doing so, the new gas pressure and density are not necessarily compatible with the observed polarization spectra from which the physical parameters were inferred. Another possibility to gain access to the geometrical scale $z$ was given by \cite{riethmueller2017}, where three-dimensional MHD simulations were used to reproduce the polarization signals of an observed area on the solar surface. Since an MHD code was employed to obtain the physical parameters, the gas pressure and density verify the MHD equation of motion in the whole of the domain employed. Moreover, physical parameters are already prescribed on the geometrical scale, and thus no $\tau\xrightarrow{}z$ conversion is needed. However, this approach requires a huge computational effort as several MHD simulations have to be run. Also, under this approach, the result of the inversion might be accurate as long as the actual physical model giving rise to the observed spectra is attained in the simulation.

Here, we put forward an alternative way to tackle the aforementioned problems. If one solves the RTE in geometrical scale, then spatial derivatives are handled directly. This strategy allows the straightforward inclusion of the Lorentz force terms in the equation of motion as well as the null divergence condition in the magnetic field to be fulfilled: $\nabla\cdot\mathbf{B}=0$. Thus it would be possible to solve the RTE self-consistently with physical constraints in three dimensions. Here we present a first step in this approach, and we implement a one-dimensional polarized RTE solver for the Zeeman regime in the geometrical scale $z$. The code is parallelized and can be used to calculate synthetic polarized spectra from MHD simulations and invert observed spectropolarimetric data in order to infer the various physical parameters (temperature, three magnetic components of the magnetic field, and LOS velocity) of the solar atmosphere. The inference of these parameters is accurate in terms of $\tau$ and depends on the ability of the user to infer the gas pressure and density in geometrical scale.

\section{Numerical model}

\subsection{Forward modeling}
\label{sec:forward_modeling}

This numerical code faces the problem of solving the RTE in a geometrical scale under the assumption of LTE (local thermodynamic equilibrium), complete redistribution approximation, and a plane parallel atmosphere. This problem is a coupled system of equations that can be expressed as:

\begin{equation}
\frac{d\,{\varmathbb I}(\lambda, s)}{d\,s}=-\hat{{\varmathbb K}}(\lambda, s)\,({\varmathbb I}(\lambda, s)-{\varmathbb S}(\lambda, s)),
\end{equation}

where $s$ is the geometrical distance along the ray path, ${\varmathbb I}(\lambda, s)$ is a wavelength-dependent pseudovector of the four Stokes parameters characterizing the polarization state of the light, $\hat{{\varmathbb K}}(\lambda, s)$ is the so-called propagation matrix, a wavelength-dependent $4\times4$ matrix that characterizes the interaction between light and the medium it travels through, and ${\varmathbb S}(\lambda, s)$ is a wavelength-dependent pseudovector for the source function. In the following, and for the sake of clarity, we avoid the explicit dependence in $\lambda$.

The derivation of the solution of the polarized RTE in the Zeeman regime has been covered extensively in the literature \citep[see e.g.,][and references therein]{landi2004} and several codes have been implemented to solve it in $\tau$-scale. Here we do not revisit those derivations. Instead, we outline the general building blocks of the formalism employed for the resolution of this problem in geometrical scale.

This equation has a formal solution in terms of a formal operator ${\varmathbb O}$ \citep{landi1985}:

\begin{equation}
{\varmathbb I}(s)\,=\,\int_{s_{0}}^{s}\,\hat{{\varmathbb O}}(s,s^{\prime})\,\hat{{\varmathbb K}}(s^{\prime})\,{\varmathbb S}(s^{\prime})\,ds^{\prime}\,+\,\hat{{\varmathbb O}}(s,s_{0})\,{\varmathbb I}(s_{0}),
\label{eq:formal_solution}
\end{equation}

where $s_{0}$ is the lower boundary, $\hat{{\varmathbb O}}(s,s_{0})$ is the Stokes-evolution operator between the lower boundary and the integration limit, and ${\varmathbb I}(s_{0})$ is the illumination from the lower boundary. This expression evaluates the leaving polarimetric properties of the light given the lower boundary and the physical properties of the medium. In this work, we consider that the medium through which light propagates is made of $N$ slabs, each of which has constant physical properties along the whole slab. Under such an assumption, ${\varmathbb O}(s,s^{\prime})$ has an analytic expression \citep[see Eq. 10,][]{landi1985}. In this case, the light leaving one of these slabs (index k) is given, following Eq. \ref{eq:formal_solution}, by
\begin{equation}
{\varmathbb I}(s_{k})\,=[1\!\!1-\hat{{\varmathbb O}}(s_{k},s_{k-1})]\,{\varmathbb S}(s_{k})\,+\hat{{\varmathbb O}}(s_{k},s_{k-1})\,{\varmathbb I}(s_{k-1}),
\label{Eq:rtelayer}
\end{equation}

where $1\!\!1$ is the $4\times4$ identity matrix, ${\varmathbb I}(s_{k})$ stands for the Stokes vector leaving slab $k$, ${\varmathbb S}(s_{k})$ is the source function in slab $k$, which is constant throughout the slab, and ${\varmathbb I}(s_{k-1})$ is the incident Stokes vector from below the slab. In the general case, we consider a set of $N$ such slabs with a proper boundary condition. In this case, it can be proven that the light leaving this medium is given by the concatenation of the various slabs:

\begin{equation}
{\varmathbb I}(s_{N})=\sum_{j=1}^{N}\hat{{\varmathbb O}}(s_{N},s_{j})\left[1\!\!1-\hat{{\varmathbb O}}(s_{j},s_{j-1})\right]{\varmathbb S}(s_{j})+\hat{{\varmathbb O}}(s_{N},s_{0}){\varmathbb I}(s_{0}).
\label{eq:code_solution}
\end{equation}

Equation \ref{eq:code_solution} expresses the light leaving the medium in terms of the physical properties of each slab (by means of the Stokes-evolution operator and the emissivity in each slab) and on the lower boundary condition ${\varmathbb I}(s_{0})$. This equation is equivalent to that presented by \citet[][Eq. 16]{deltoroiniesta1995}.

For the boundary condition ${\varmathbb I}(s_{0})$, we note that in a deep-enough layer the temperature, the gas pressure, and the density are high enough to cause the opacity to be sufficiently large so as to ensure that the Stokes-evolution operator becomes negligible. Therefore, from Eq. \ref{Eq:rtelayer}, if the Stokes-evolution operator is small enough, from that deep-enough layer, the leaving beam of such a deep layer depends only in its own physical properties through the source function ${\varmathbb I}(s_{0})={\varmathbb S}(s_{0})$. As we are assuming LTE, ${\varmathbb S}(s_{k})=(B(T(s_{k})),0,0,0)^{\dagger}$, where $B(T(s_{k}))$ is the Planck function for the temperature at $s_{k}$.

With this formalism one can solve the forward problem for the RTE. To do so, one needs to specify some additional features:

\begin{itemize}
\item An equation of state (EOS): This is set to be given by an ideal gas equation with a variable molecular weight to account for free electrons.
\item Continuum opacities: We follow a reduced scheme of the so-called NICOLE opacity package \citep[see Sect. 4.1.3 in][]{socasnavarro2015}, excluding the contributions from Rayleigh scattering on molecular hydrogen and the contributions of bound-free and free-free terms due to magnesium.
\item Line opacity: This is calculated following \cite{landi1976}, excluding the non-LTE departure coefficients as here we restrict ourselves to the LTE case.
\item Partition functions: These are implemented following \cite{wittmann1974} who allowed analytical expressions for both the partition functions and their derivatives with respect to the temperature. This set of functions includes 92 species and 3 ionization states. We note that for the hydrogen atom, the first ionization stage involves the negative hydrogen.
\item Line broadening due to collisions: Two different treatments are available: 1) The ABO theory \citep{anstee1995,barklem1998}, which requires additional parametric information to estimate the collisional broadening, namely the cross section ($\sigma$) and the velocity parameter ($\alpha$). These can be supplied or they can be internally calculated (provided the energy limit for each energy level involved in the transition is known) from the tabulated tables \citep{anstee1995,barklem1997,barklem1998}. 2) In the absence of the aforementioned data, the classic Uns\"old approximation \citep{unsold1955} is used. In this case, the line broadening is calculated without any additional atomic parameters. However, this formalism is known to underestimate the collisional broadening giving rise to narrower spectral lines \citep{barklem1998c}. 
\item Magnetic field effect: At this point the code takes into account only the Zeeman effect for the Stokes parameters, following the formalism of \cite{landi1976} but changing the sign of Stokes V so that the code possesses the same sign criteria for the polarity of magnetic field as SIR, SPINOR, or NICOLE.
\item Spectral line blends are treated self-consistently.
\end{itemize}

With these ingredients one is able to perform the forward modeling for any solar or stellar atmosphere prescribed in $N$ slabs in geometrical scale. Furthermore, given this parametric model in geometrical scale, it is possible to face the inverse solver: namely,  to infer the various physical parameters as a function of the geometrical scale given a set of spectropolarimetric observations. To the best of our knowledge, this is the first code capable of doing so. It is worth emphasizing that this RTE solver, as prescribed here, has analytical expressions for all the model parameters. This is extremely advantageous because it allows one to obtain analytical derivatives of the spectropolarimetric spectra (i.e., Stokes pseudovector) with respect to the various physical properties of the medium. These are the so-called response functions, and are of critical importance as they are usually required in the resolution of the inverse problem.

\subsection{Response functions}
\label{sec:responsefunctions}

Response functions give the variation of the leaving Stokes pseudovector, ${\varmathbb I}(s_{N}),$ due to the variation of any physical parameter $x$ at a given height $s_{k}$ (referred to as $x_{k}=x(s_{k})$). The various $x_{k}$ refer to the temperature ($T$), the gas pressure ($P_{g}$), the density ($\rho$), the line-of-sight velocity ($v_{LOS}$), and the three cartesian components of the magnetic field ($B_{x}$, $B_{y}$, and $B_{z}$) at the $k$-layer of the atmospheric model. In our case, taking derivatives in Eq. \ref{eq:code_solution}:

\begin{equation}
\begin{split}
\frac{\partial}{\partial\,x_{k}}{\varmathbb I}(s_{N})&=\frac{\partial}{\partial\,x_{k}}\Bigg\{\sum_{j=1}^{N}\hat{{\varmathbb O}}(s_{N},s_{j})\left[1\!\!1-\hat{{\varmathbb O}}(s_{j},s_{j-1})\right]{\varmathbb S}(s_{j})+\\
&\qquad\qquad\hat{{\varmathbb O}}(s_{N},s_{0}){\varmathbb I}(s_{0})\Bigg\}.
\end{split}
\end{equation}

We now take into account that the only parameters that depend on $x_{k}$ are $\hat{{\varmathbb O}}(s_{k},s_{k-1})$ and ${\varmathbb S}(s_{k})$, and so it can be shown that

\begin{equation}
\begin{split}
\frac{\partial}{\partial\,x_{k}}{\varmathbb I}(s_{N})&=\hat{{\varmathbb O}}(s_{N},s_{k})\Bigg[\\
&\delta_{k}\hat{{\varmathbb O}}(s_{k},s_{k-1})\sum_{j=1}^{k-1}\hat{{\varmathbb O}}(s_{k-1},s_{j})\,\left[1\!\!1-\hat{{\varmathbb O}}(s_{j},s_{j-1})\right]{\varmathbb S}(s_{j})\\
&-\delta_{k}\hat{{\varmathbb O}}(s_{k},s_{k-1}){\varmathbb S}(s_{k})+\left[1\!\!1-\hat{{\varmathbb O}}(s_{k},s_{k-1})\right]\delta_{k}{\varmathbb S}(s_{k})\\
&+\delta_{k}\hat{{\varmathbb O}}(s_{k},s_{k-1})\hat{{\varmathbb O}}(s_{k-1},s_{0})\,{\varmathbb I}(s_{0})
\Bigg],
\end{split}
\label{Eq:responsefunction_generalexpression}
\end{equation}

where $\delta_{k}$ stands for $\frac{\partial}{\partial x_{k}}$. As mentioned before, under the assumption of LTE, the source function at $s_{k}$, ${\varmathbb S}(s_{k})$, corresponds to the black-body emission and depends only on the temperature at that height $T_{k}=T(s_{k})$. Therefore, derivatives of ${\varmathbb S}(s_{k})$ with respect to any other physical parameter other than the temperature vanish.  The derivatives of the Stokes-evolution operator with respect to the physical parameters can be determined by applying the chain rule to the expressions provided by \cite[Eq. 10]{landi1985}. It can be shown that these derivatives depend only on the derivatives of the various elements of the propagation matrix $\hat{{\varmathbb K}}(s)$: $\eta_{I,Q,U,V}(s)$ and $\rho_{Q,U,V}(s)$. For the sake of clarity and completeness we present them here in terms of the cartesian components of the magnetic field:

\begin{equation}
\begin{split}
\eta_{I}&=\kappa_{c}+\frac{\kappa_{l}}{2}\Bigg[\phi_{p}\Bigg(1-\frac{B_{z}^{2}}{B^{2}}\Bigg)+\frac{1}{2}(\phi_{b}+\phi_{r})\Bigg(1+\frac{B_{z}^{2}}{B^{2}}\Bigg)\Bigg],\\
\eta_{Q}&=\frac{\kappa_{l}}{2}\Bigg[\phi_{p}-\frac{1}{2}(\phi_{b}+\phi_{r})\Bigg]\Bigg(1-\frac{B_{z}^{2}}{B^{2}}\Bigg)\frac{B_{x}^{2}-B_{y}^{2}}{B_{x}^{2}+B_{y}^{2}},\\
\eta_{U}&=\frac{\kappa_{l}}{2}\Bigg[\phi_{p}-\frac{1}{2}(\phi_{b}+\phi_{r})\Bigg]\Bigg(1-\frac{B_{z}^{2}}{B^{2}}\Bigg)\frac{2\,B_{x}\,B_{y}}{B_{x}^{2}+B_{y}^{2}},\\
\eta_{V}&=\frac{\kappa_{l}}{2}(\phi_{r}-\phi_{b})\frac{B_{z}}{B},\\
\rho_{Q}&=\frac{\kappa_{l}}{2}\Bigg[\psi_{p}-\frac{1}{2}(\psi_{b}+\psi_{r})\Bigg]\Bigg(1-\frac{B_{z}^{2}}{B^{2}}\Bigg)\frac{B_{x}^{2}-B_{y}^{2}}{B_{x}^{2}+B_{y}^{2}},\\
\rho_{U}&=\frac{\kappa_{l}}{2}\Bigg[\psi_{p}-\frac{1}{2}(\psi_{b}+\psi_{r})\Bigg]\Bigg(1-\frac{B_{z}^{2}}{B^{2}}\Bigg)\frac{2\,B_{x}\,B_{y}}{B_{x}^{2}+B_{y}^{2}},\\
\rho_{V}&=\frac{\kappa_{l}}{2}(\psi_{r}-\psi_{b})\frac{B_{z}}{B},
\end{split}
\label{eq:propagation_matrix_elements}
\end{equation}

where $\kappa_{c}$ and $\kappa_{l}$ are the continuum and line opacity, respectively, $B$ is the magnetic field modulus, and the various $\phi_{b,p,r}$ and $\psi_{b,p,r}$ are the absorption and dispersion profiles, respectively, whose formulation can be found elsewhere \cite[see e.g., Ch. 5 of ][]{landi2004}.

From these equations, it can be shown that the derivatives of the various expressions in Eq. \ref{eq:propagation_matrix_elements} with respect to $v_{LOS}$, $B_{x}$, $B_{y}$, and $B_{z}$ are straightforward by means of the application of the chain rule and taking into account that $\kappa_{c}$ and $\kappa_{l}$ do not depend on these physical parameters. The derivatives involving thermodynamic properties, however, require additional consideration.

\subsubsection{Thermodynamic derivatives of the propagation matrix elements}
\label{Sec:therder}
Here we focus on the calculation of the partial derivatives of the Stokes pseudovector with respect to the thermodynamic parameters: temperature, $T$, gas pressure, $P_{g}$ and density $\rho$. These derivatives have to take into account that there is an additional constraint (EOS) that relates them to one another. Therefore, in contrast to the derivatives with respect to the line-of-sight velocity, $v_{LOS}$, and the cartesian components of the magnetic field vector, $B_{x}$, $B_{y}$, and $B_{z}$, where all remaining parameters are assumed constant, for the derivatives with respect to  $T$, $P_{g}$, and $\rho$, there are always two nonconstant parameters. For instance, if we have the derivative with respect to $T$, and we assume that $\rho$ is constant; subsequently, $P_{g}$ must change so that the EOS is fulfilled. For this reason, derivatives with respect to $T$, $P_{g}$, or $\rho$ explicitly state which of them is assumed constant, that is, the Stokes derivative with respect to $T$ at constant $\rho$ is expressed as: $\left(\frac{\partial {\varmathbb I}(s_{N})}{\partial T}\right)_{\rho}$. Therefore, here we are interested in obtaining

\begin{equation}
    \left(\frac{\partial{\varmathbb I}(s_{N})}{\partial\,T(k)}\right)_{\rho},
    \left(\frac{\partial{\varmathbb I}(s_{N})}{\partial\,P_{g}(k)}\right)_{T},
    \left(\frac{\partial{\varmathbb I}(s_{N})}{\partial\,\rho(k)}\right)_{T}
    \label{Eq:stokesvectorderivatives}
.\end{equation}

The selection of the thermodynamical parameter to be kept constant during the inversion is an arbitrary decision as they give equivalent results. It is important to bear in mind however that the computation of $\left(\frac{\partial{\varmathbb I}(s_{N})}{\partial\,P_{g}(k)}\right)_{T}$ is computationally cheaper than that of $\left(\frac{\partial{\varmathbb I}(s_{N})}{\partial\,P_{g}(k)}\right)_{\rho}$, as the former does not involve changes in temperature, which is by far the most frequent parameter in the equations to be differentiated. For the derivatives with respect to T, there is no difference in using either $\left(\frac{\partial{\varmathbb I}(s_{N})}{\partial\,T(k)}\right)_{\rho}$ or $\left(\frac{\partial{\varmathbb I}(s_{N})}{\partial\,T(k)}\right)_{P_{g}}$.

For the partial derivative $\left(\frac{\partial{\varmathbb I}(s_{N})}{\partial\,T(k)}\right)_{\rho}$, we need the calculation of $\delta_{k}{\varmathbb S}(s_{k})$ in Eq.~\ref{Eq:responsefunction_generalexpression}. In addition, the partial derivatives with respect to the thermodynamic parameters in Eq.~\ref{Eq:stokesvectorderivatives}, require the calculation of the derivatives of the Stokes evolution operator: $\delta_{k}\hat{{\varmathbb O}}(s_{k},s_{k-1})$ in Eq.~\ref{Eq:responsefunction_generalexpression}. At the same time, these derivatives of the folloing Stokes evolution operator rely upon the calculation of $\delta_{k}\left\{\eta_{I},\eta_{Q},\eta_{U},\eta_{V},\rho_{Q},\eta_{U},\eta_{V}\right\}$. In this particular case, these involve the determination of the derivatives of the continuum $\kappa_c$ and line $\kappa_{l}$ absorption coefficients, which in turn depend on the number of electrons per cubic centimeter (i.e., electron density) $n_e$. Therefore, after applying the chain rule over the Stokes evolution operator, what we need to evaluate is\\

\begin{equation}
    \left(\frac{\partial n_{e}}{\partial T}\right)_{\rho},
    \left(\frac{\partial n_{e}}{\partial P_{g}}\right)_{T},
    \left(\frac{\partial n_{e}}{\partial \rho}\right)_{T}
    \label{Eq:electrondensityderivatives}
.\end{equation}

These derivatives require additional algebra, yet they can be analytically calculated, as is shown in Appendices \ref{App:dnedt_rho},  \ref{App:dnedpg_t}, and \ref{App:dnedrho_t}.

\subsection{Inverse solver}
\label{sec:inverse_solver}

Up to here, we have faced the problem of obtaining the polarized spectra provided a set of height-dependent physical stratifications ${\varmathbb I}(\lambda_{i})^{\text{syn}}=f(\lambda_{i}, \mathbf{X})$, where $\lambda_{i}$ are the independent variables (the number of wavelength times the number of Stokes profiles observed, $n_{\nu}$) and $\mathbf{X}$ the set of free parameters defining the forward model. It is sometimes interesting to proceed the other way around, that is, one records the polarized spectra from an object and aims at inferring some meaningful physical properties of the plasma from where the light comes from. This is called the inverse problem, where one looks for the model parameters that best fit, under some approximations, the observations ${\varmathbb I}(\lambda_{i})^{\text{obs}}$ by defining a merit function:

\begin{equation}
\mathbf{X}^{\prime}\equiv \text{arg}\,\text{min}{\frac{1}{n_{\nu}-n_{f}}\sum_{i}^{n_{\nu}}\left[\frac{f(\lambda_{i}, \mathbf{X})-{\varmathbb I}(\lambda_{i})^{\text{obs}}}{\sigma_{i}}\right]^2},
\end{equation}

 where $n_{f}$ stands for the number of free parameters and $\sigma_{i}$ for the uncertainty and/or weighting of each observed data point. There are several strategies and algorithms when facing such an optimization problem, both local and globally. In this work we have chosen to implement a LM (Levenberg-Marquardt,  \citealt{levenberg1944} and \citealt{marquadt1963}) algorithm to perform this step. This kind of algorithm starts from a provided guess of the set of model parameters ($\mathbf{X}$) and looks for the optimization by iteratively modifying these parameters. To do so, the perturbations of the free parameters are evaluated following

\begin{equation}
{\boldsymbol \delta}=({\mathbf H}+\Lambda\, \text{diag}({\mathbf H}))^{-1}\,{\mathbf J(\lambda)}^T(f(\lambda, \mathbf{X})-{\varmathbb I}(\lambda)^{\text{obs}}), 
\label{Eq:perturbation}
\end{equation}

where ${\boldsymbol \delta}$ is the perturbation of the set of the model free parameters, $\mathbf{H}\approx{\mathbf J(\lambda)}^T\,{\mathbf J(\lambda)}$ is an approximation to the so-called Hessian matrix, ${\mathbf J(\lambda)}$ is the Jacobian matrix, and $\Lambda$ is the dampening factor. Here it is worth emphasizing that the calculation of the perturbation requires the Jacobian matrix, which in our code is calculated analytically at the same time as the forward problem is solved by means of the response functions described in Sect. \ref{sec:responsefunctions}. It is important to note that the Jacobian makes use of dimensionless response functions, which are calculated by multiplying $\frac{\partial}{\partial\,x_{k}}{\varmathbb I}(s_{N})$ in Eq. \ref{Eq:responsefunction_generalexpression} by a typical value of the physical parameter. We return to this particular point in Sect. \ref{Sec:fwd_mdl}. This point speeds up the calculations since otherwise they have to be numerically evaluated solving the forward model additional times for each iteration of the LM algorithm. Each time a perturbation, $\boldsymbol{\delta}$, to the model parameters is evaluated, the objective function of these new parameters is checked as a ruler to approve or reject the proposed perturbation.

An important point to notice here is that the first term on the right-hand-side of Eq. \ref{Eq:perturbation} may involve (as it does for instance in the examples presented in the following section) the inversion of large and quasi-singular matrices. In order to tackle this point we follow the strategy introduced by \cite{ruizcobo1992} and use the Singular Value Decomposition (SVD) method\footnote{The SVD is computed using Linear Algebra PACKage (LAPACK).}. This way, it is possible to solve an ill-posed problem by reducing its dimensions. This reduction is controlled by the number of eigenvalues that one keeps when calculating the inverse dampened-Hessian matrix. In this fashion, it is possible to find the optimum solution, in terms of least-squares, for that number of eigenvalues. The choice of the threshold value is, for Stokes inversion codes, somewhere between $10^{-2}$ and $10^{-6}$ times the largest eigenvalue. In all examples in this paper we have used $10^{-4}$.

The calculation on the uncertainties in the inference of the various $\mathbf{X}^{\prime}$ is performed assuming that each free parameter is independent of the others. In such a case, the covariance matrix is diagonal and given by the inverse of the diagonal elements of the Hessian. Therefore, the uncertainties are given by:

\begin{equation}
  \Delta\mathbf{X}^{\prime}=\sqrt{\frac{\chi^2}{\text{diag}(\mathbf{H})}}, 
\end{equation}

where $\chi^2=\frac{1}{n_{\nu}-n_{f}}\sum_{i}^{n_{\nu}}\left[\frac{f(\lambda_{i}, \mathbf{X})-{\varmathbb I}(\lambda_{i})^{\text{obs}}}{\sigma_{i}}\right]^2$. Since the uncertainties calculated this way neglect possible correlation between the various free parameters, the result is a lower limit for the uncertainties of each inferred parameter.

In principle, one can invert the whole set of physical parameters defining the atmosphere at once. However, this strategy can lead to inversion results that might not actually fit the profiles or at least, not as well as one would expect. This is because the minimization algorithm might get stuck in a local minimum which is not as deep as others might be. This is usually the case if one is not fairly close to the actual minimum and there are several free parameters (for instance 256 or 1280 as below). Instead, as introduced by \cite{ruizcobo1992}, it is better to start first with a reduced problem where the number of free parameters is much lower than in the original problem and is then progressively increased as one approaches the minimum for that reduced problem. For this reason, we perform the fitting process in a sequential manner: we progressively increase the number of equivalent slabs perturbed, $\boldsymbol{\delta}$, from a constant in height perturbation for each parameter, until all the slabs are perturbed individually. Since response functions are given for all heights, we employ the concept of equivalent response functions \citep[see e.g., Appendix A of][]{ruizcobo1992} to linearly interpolate from the inversion grid into the model height grid at each iteration step.

\subsection{Parallelization}

\begin{figure*}
\centering
\includegraphics[width=\hsize]{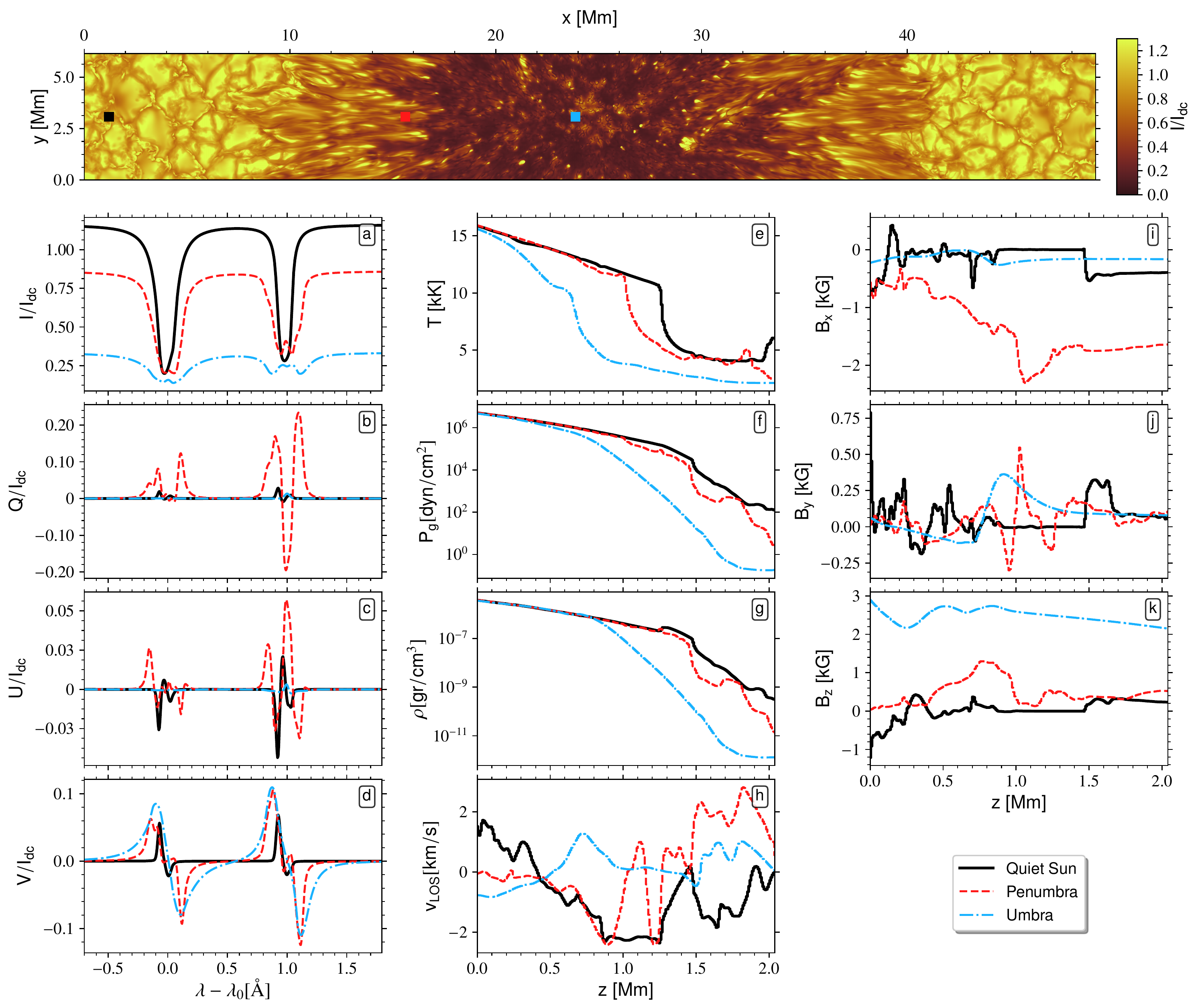}
   \caption{Examples of three simulated Stokes profiles as RTE is solved for a snapshot of the Rempel simulation of a sunspot \citep{rempel2012}. Top panel: Monochromatic continuum intensity map in which three pixels are highlighted in black, red, and light-blue squares as representatives of low magnetized, penumbral, and umbral areas, respectively; their spectral profiles are presented following this same color code in panels from (a) to (d) for Stokes I, Q, U, and V, respectively. The physical parameters ($T(z)$, $P_{g}(z)$, $\rho(z)$, $v_{LOS}(z)$, $B_{x}(z)$, $B_{y}(z)$, and $B_{z}(z)$) that lead to these profiles are shown in panels (e) to (k), respectively.
           }
      \label{Fig:ProfSyn}
\end{figure*}

Modern solar observing facilities such as Hinode \citep{kosugi2007}, the Solar Dynamics Observatory \citep{scherrer2012}, the Swedish Solar Telescope \citep{scharmer2003}, the Goode Solar Telescope \citep{cao2010}, and GREGOR \citep{schmidt2012} provide large amounts of data that currently can barely be processed at the needed speed \citep{borrero2011}. In the future this issue will be aggravated with the arrival of new observatories such as the Daniel K.~Inouye Solar Telescope (DKIST; \citealt{rimmele2010}) and the European Solar Telescope (EST; \citealt{collados2010}) whose data throughput will be several orders of magnitude larger than that of the current facilities. For this reason, it is currently compulsory to build new forward or inverse solvers for the RTE that are parallelized. Moreover, it is nowadays common practice for theoreticians and modelers, in particular developers of three-dimensional magnetohydrodynamic codes such as MURAM \citep{rempel2009}, CO5BOLD \citep{freytag2012}, and MANCHA \citep{gonzalez2018}, and even observers, to employ forward solvers for the polarized RTE to produce simulated observations from snapshots of the MHD simulations \citep{danilovic2008,danilovic2010a,danilovic2010b,borrero2010,borrero2014,lites2017} in order to compare with real observations and to make predictions, so as to study under what circumstances a particular feature seen in the simulations could be detected observationally. The large number of available snapshots and the ever increasing size of the simulation boxes call also for fast (i.e., parallel) codes to solve the RTE.\\

\begin{figure*}
\centering
\includegraphics[width=\hsize]{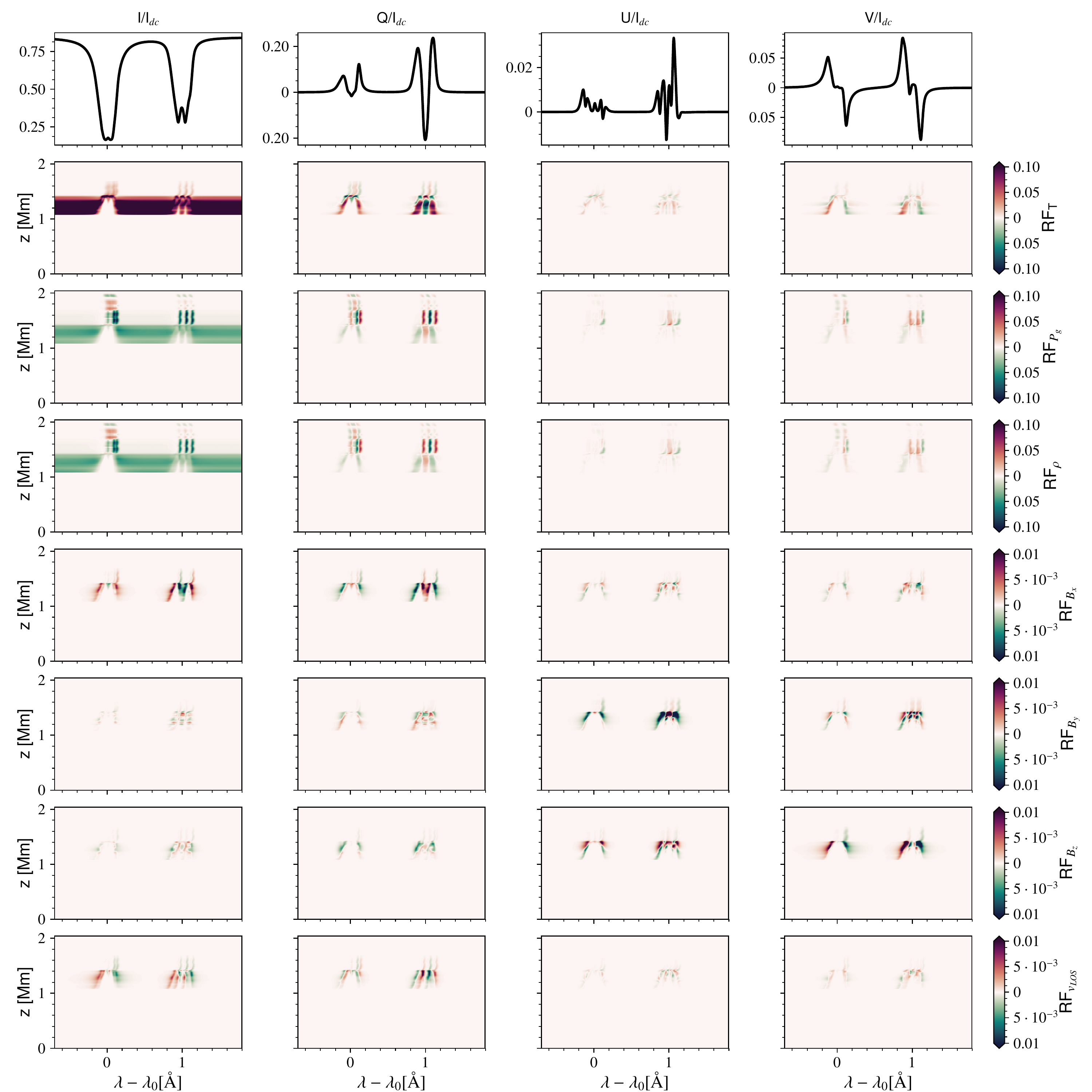}
   \caption{Analytical dimensionless response functions for, from the second row to the last one, $T(z)$, $P_{g}(z)$, $\rho(z)$, $B_{x}(z)$, $B_{y}(z)$, $B_{z}(z)$, and $v_{LOS}(z)$ for the penumbral pixel highlighted in Fig. \ref{Fig:ProfSyn} (red square) and whose profiles are shown on the first row. Columns, from left to right: Stokes I, Q, U, and V.}
      \label{Fig:ProfSynRF}
\end{figure*}

\begin{figure*}
\centering
\includegraphics[width=\hsize]{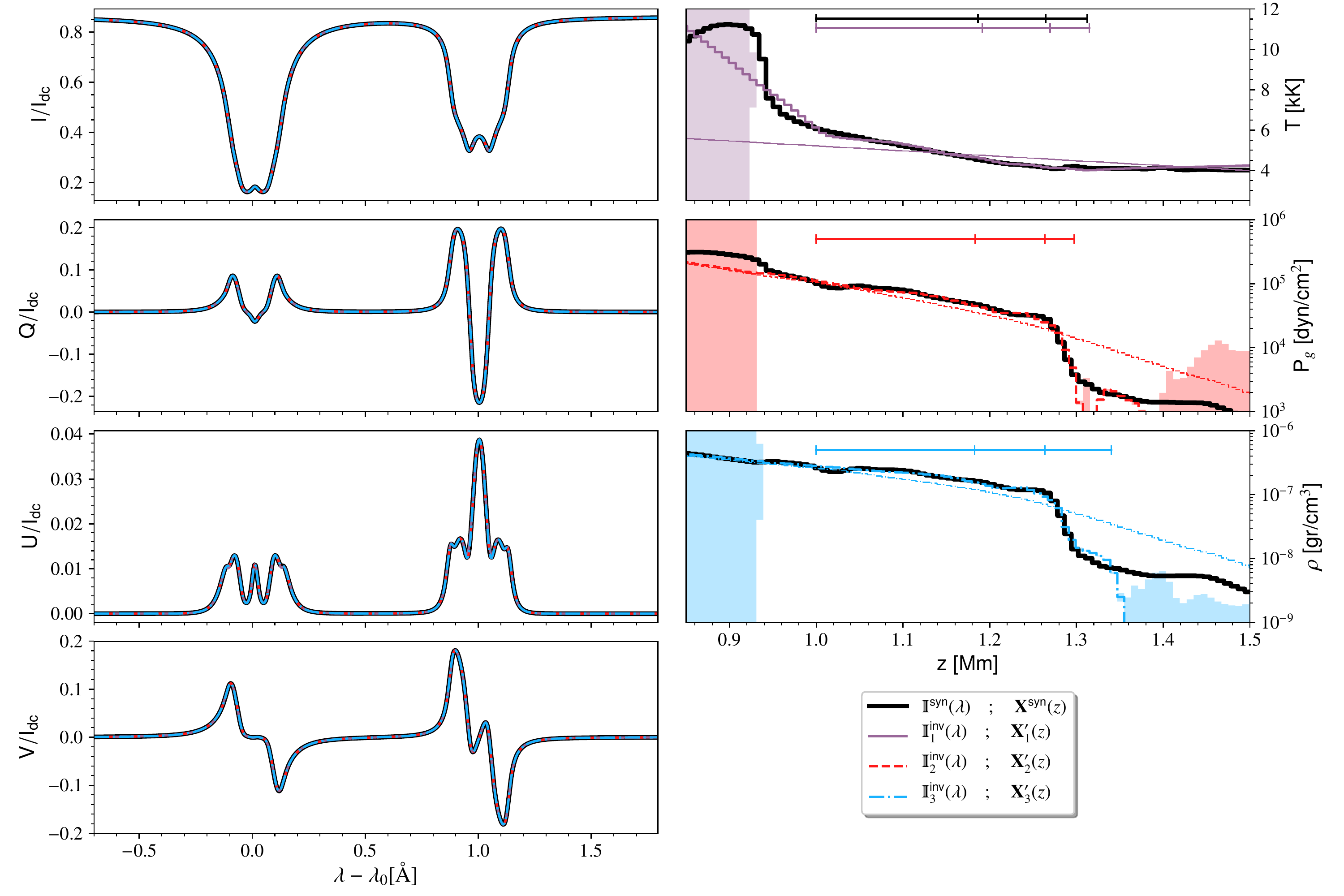}
   \caption{Results of the inversion of three thermodynamic parameters. Three different inversions are presented, in each of which a single physical parameter is inverted: the temperature, $T(z)$, (solid-purple line), the gas pressure, $P_{g}(z)$, (dashed-red), and the density, $\rho(z)$, (dash-dotted-blue). Left panels: Simulated spectral lines inverted (${\varmathbb I}^{\text{syn}}(\lambda)$, in black) coming from the same penumbral pixel highlighted in Fig. \ref{Fig:ProfSyn} (red square) together with the results of the three inversions. Right panels: Thermodynamic parameters (synthetic $\mathbf{X}^{\text{syn}}(z)$ and inverted $\mathbf{X}^{\prime}(z)$), from top to bottom, $T(z)$, $P_{g}(z)$, and $\rho(z),$ (the initialization for each inversion is shown in thin lines for each thermodynamical parameter). Line-of-sight velocity and magnetic field components are taken to be constant with height. For the sake of clarity, in each panel we only show the physical parameter inverted, as the other ones are taken as known. The colored shadows represent the uncertainties as derived by the code (see Sect. \ref{sec:inverse_solver}). Right panels include a horizontal line for each physical parameter (following the same color code) with ticks for, from left to right, the heights at which \mbox{$\lg\tau_{5}=0,-1,-2,-3$}.}
      \label{Fig:InvTheSin}
\end{figure*}

With this in mind the forward and inverse code presented in this work has been conceived, from its initial stages, to being fully parallelized. The implementation is straightforward as each pixel on the observed solar surface ($x$, $y$) is treated independently and requires no information from neighboring pixels. Hence, we have chosen a master-slave parallelization scheme where the master process reads, distributes the work, and writes the outputs. Slave processes are used to solve the RTE and/or solve the inverse problem for each pixel independently.

\section{Results}

\subsection{Forward modeling}
\label{Sec:fwd_mdl}

Given the implementation detailed in the previous section, one can solve the RTE for polarized light under LTE and in the Zeeman regime for any atmospheric stratification. In Fig. \ref{Fig:ProfSyn} (top panel) we show the continuum intensity of a snapshot from the \cite{rempel2012} sunspot simulation. The size of the simulation box is $4096\times512\times256$ with a grid size of $\Delta x=10$, $\Delta y=10$, and $\Delta z=8$ km. We additionally (remaining panels in Fig. 1) show three different simulated Stokes profiles, ${\varmathbb I}^{\text{syn}}(\lambda)$, together with their physical parameter stratification chosen from this snapshot. The synthetic profiles here, and all the forthcoming tests in this paper, are synthesised using the same spectral lines and wavelength grid. We used the two widely used Fe{\sc I} spectral lines at $6301.5$ and $6302.5$ {\AA}. The wavelength grid runs from -700 m{\AA} to +1800 m{\AA} with respect to the 6301.5 {\AA} spectral line with a step of 5 m{\AA}. This setup is chosen so that the simultaneous inversion of the temperature, LOS velocity, and the three cartesian components of the magnetic field can be performed using the original height grid of the MHD simulation (256 heights). In a more realistic case, one can limit the number of free parameters setting the maximum number of heights at which perturbations of the model are calculated. Subsequently, all the intermediate heights are linearly interpolated from the previous ones.

One can also retrieve the response functions for a supplied atmosphere. As stated in Sect. \ref{sec:responsefunctions}, these functions encode information about where Stokes profiles are sensitive, to what they are sensitive, and to what degree. To the best of our knowledge, this is the first code capable of calculating response functions analytically at geometrical scale height. As an example, Fig. \ref{Fig:ProfSynRF} shows the Stokes profiles (top row) in a penumbral pixel along with the corresponding dimensionless response functions to temperature $T$ (second row), gas pressure $P_{g}$ (third row), and density $\rho$ (fourth row), three components of the magnetic field $B_{x}$, $B_{y}$, and $B_{z}$ (fifth to seventh rows, respectively), and $v_{LOS}$ (eighth row). The normalization factors employed to make the response functions dimensionless are: $\bar{T}=6000$ K, $\bar{\rho}=3\times10^{-7}$ gr cm$^{3}$, $\bar{P_{g}}=10^{5}$ dyn cm$^{-2}$, $\bar{B_{x}}=1000$ G (likewise for the other components of the magnetic field), and $\bar{v_{LOS}}=10^{5}$ cm s$^{-1}$. It is important to emphasize that the response functions to gas pressure and density are not negligible compared with the response functions to other physical parameters. The high sensitivity to these two parameters could be surprising, as when solving the RTE in optical depth scale, the Stokes profiles are only slightly sensitive to them. The difference resides in the fact that solving the RTE in geometrical scale (as is the case here), the optical depth scale is affected by the combination of temperature, gas pressure, and density. Consequently, the thermodynamic response functions allow for a better insight into how (i.e., what combination of thermodynamic parameters) makes the optical depth scale behave as it does. For instance, far away from the line core of the two spectral lines in Fig. \ref{Fig:ProfSynRF}, the response functions of Stokes I to $P_{g}$ and $\rho$ are not zero and behave more like the response functions to the temperature and unlike the response function to $B_{x}$, $B_{x}$, $B_{x}$, and so on. Consequently, the gas pressure and density have an important effect on the continuum intensity by changing the geometrical height $z$ at which the continuum level $\tau_{5}=1$ is formed.

\subsection{Inverse solver}
\label{sec:inverse_solver_results}

\begin{figure*}
\centering
\includegraphics[width=\hsize]{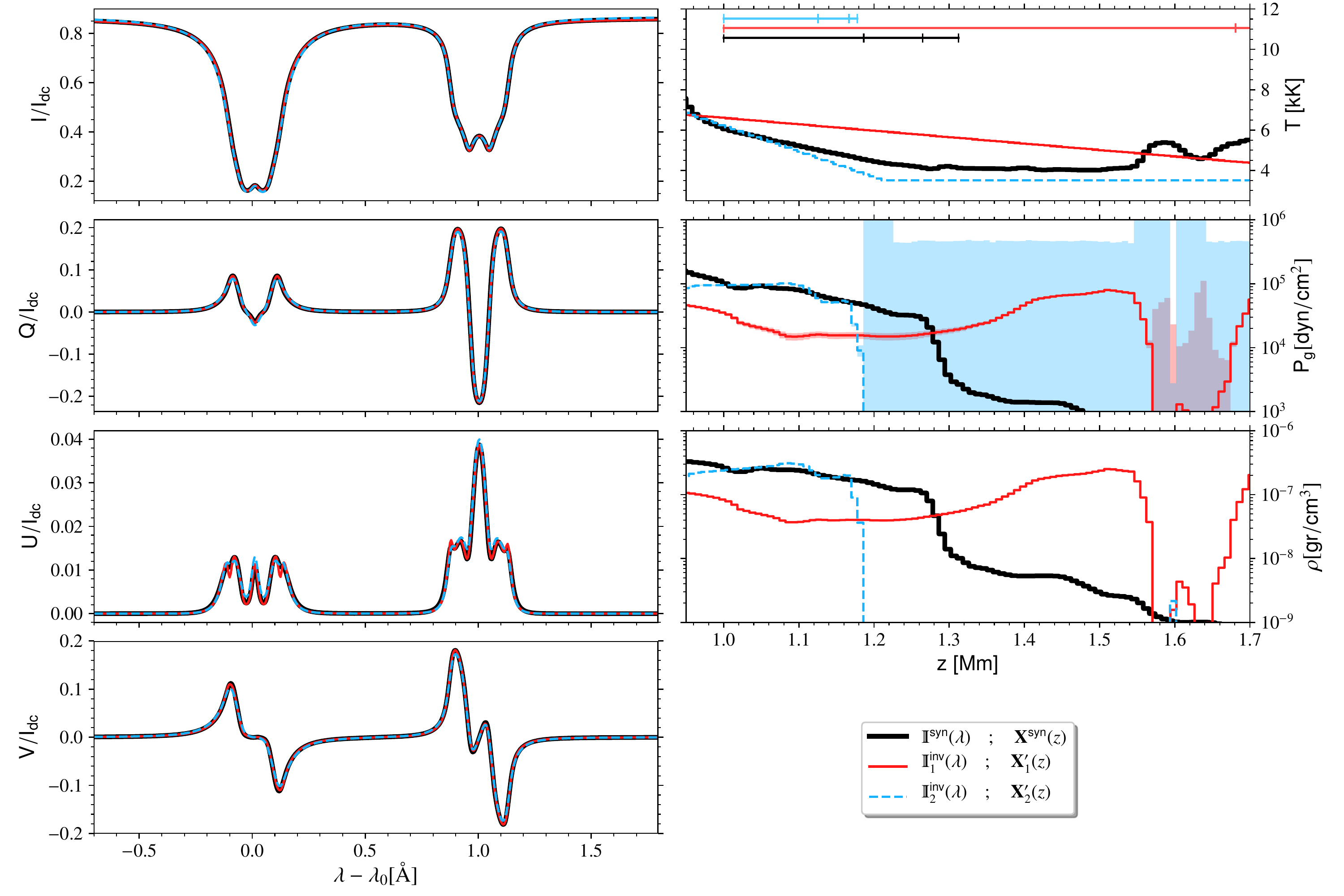}
   \caption{Two different inversions of the gas pressure, $P_{g}(z)$, for the penumbral pixel highlighted in Fig. \ref{Fig:ProfSyn} by a red square. Left column, from top to bottom: Stokes I, Q, U, and V for the simulated case (${\varmathbb I}^{\text{syn}}(\lambda)$ solid-black lines) and for the two inversions (${\varmathbb I}^{\text{inv}}_{1}(\lambda)$ solid-red lines and ${\varmathbb I}^{\text{inv}}_{2}(\lambda)$ dashed-blue lines) with fixed and `different from the simulated case' temperatures: $T_{1}(z)$ and $T_{2}(z)$, respectively. Right column: Thermodynamic parameters (synthetic $\mathbf{X}^{\text{syn}}(z)$ and inverted $\mathbf{X}^{\prime}(z)$) of the various models, following the same color code. The upper panel includes a horizontal line for each model atmosphere (following the same color code) with ticks for, from left to right, the heights at which \mbox{$\lg\tau_{5}=0,-1,-2,-3$}.
           }
      \label{Fig:ProfSynAmb}
\end{figure*}

The inverse solver aims at inferring the vertical stratification of the various physical parameters of the atmosphere $\mathbf{X}(z)$, provided a set of observed Stokes profiles, ${\varmathbb I}(\lambda)^{\text{obs}}$. To test the performance of our inverse solver we take as observed Stokes profiles the simulated ones ${\varmathbb I}^{\text{syn}}(\lambda)$ of the spectral lines described in Sect. \ref{Sec:fwd_mdl}, in particular the penumbral pixel (red case in Fig. \ref{Fig:ProfSyn}). The number of data points is 2000 (500 wavelength points for each of the four Stokes parameters). In principle, the inference can be done for any combination of: temperature, $T(z)$, gas pressure, $P_{g}(z)$, density, $\rho(z)$, the three cartesian components of the magnetic field, $B_{x,y,z}(z)$ and the LOS velocity, $v_{LOS}(z)$. The only restriction is that it is not possible to simultaneously invert all three thermodynamic parameters: $T(z)$, $P_{g}(z)$ and $\rho(z)$.  This occurs because provided two of them, the third one can be determined through the EOS. Therefore, inverting the three of them simultaneously can potentially lead to thermodynamic physical parameters that are not consistent with the EOS.

In this paper we focus mostly on the influence that the thermodynamic parameters have on an inversion code that works in the geometrical scale $z$ instead of the usual optical-depth $\tau$. Because of this, most of the forthcoming discussion avoids the performance of the code in the determination of the three components of the magnetic field and LOS velocity.

As a first test we only invert one of the three thermodynamic physical parameters at a time. The other two thermodynamic parameters as well as the three components of the magnetic field and $v_{LOS}$ are not inverted at this point. In addition, and this is the case for all the forthcoming tests but the last, we set the three components of the magnetic field and $v_{LOS}$ to a constant in height value, meaning they do not encode any information about the height. The inverted parameters are allowed to change at each of the 256 vertical heights (same as the MHD simulation employed to produce ${\varmathbb I}^{\text{syn}}(\lambda)$) and therefore in this test we have 256 free parameters. 

In Fig. \ref{Fig:InvTheSin} we depict the real (black solid lines), starting guess (thin color lines), and inferred (solid color lines) stratifications for the temperature $T(z)$ (purple), gas pressure $P_{g}(z)$ (red), and density $\rho(z)$ (blue) for the three different inversions. As can be seen, the recovered parameters are in good agreement in the region where the profiles are sensitive to: $z\in[1000, 1300]$ km as obtained from the heights at which $z(\lg\tau_{5}=0)\approx1000$ and $z(\lg\tau_{5}=-3)\approx1300$ km (see horizontal color bars in this plot). Shadowed areas represent the lower limit of the uncertainties on the inference of each physical parameter. These error bars are tiny because of the absence of noise in the profiles together with the large number of data points considered.

\begin{figure*}
\centering
\includegraphics[width=\hsize]{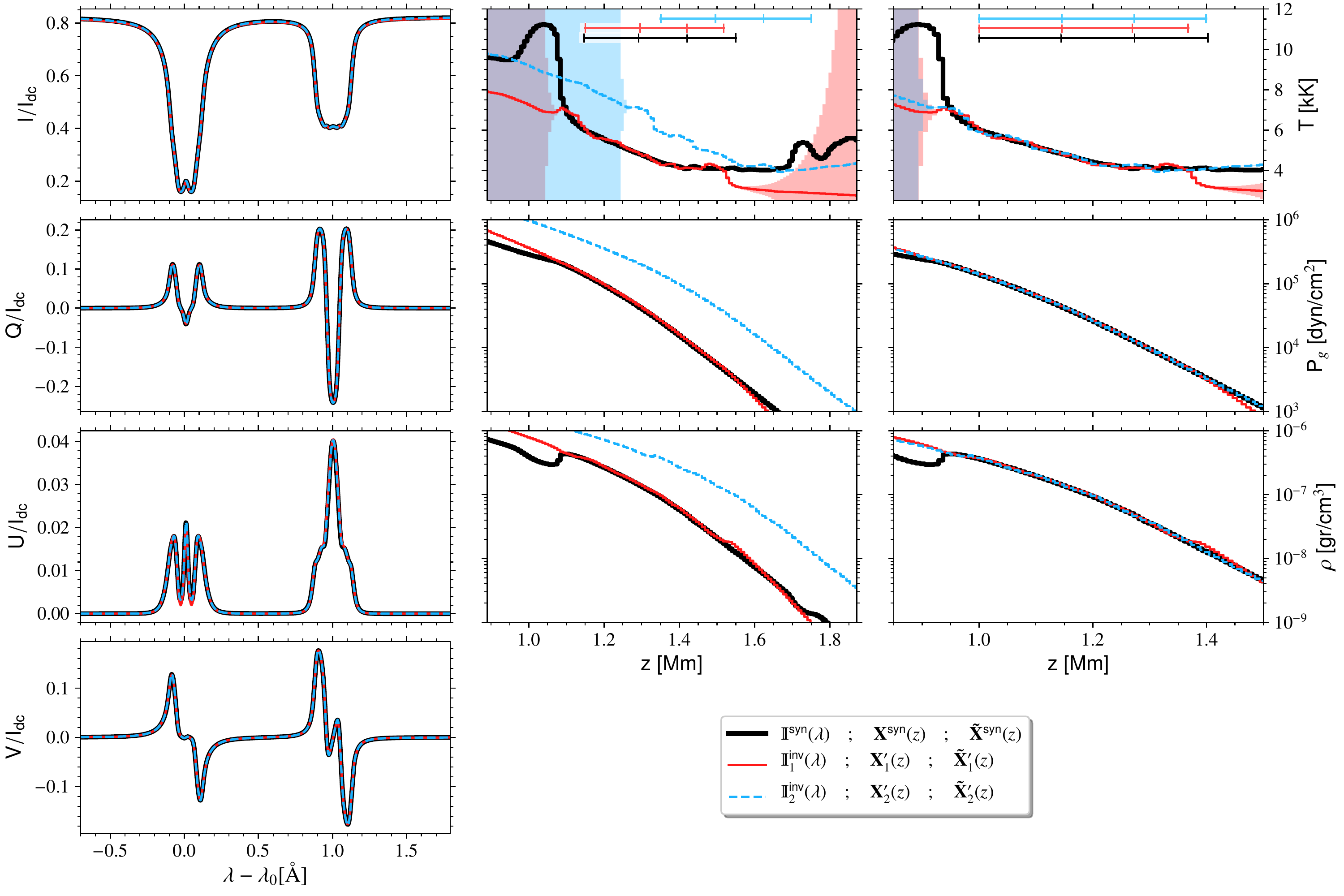}
   \caption{Inversion of Stokes profiles in hydrostatic equilibrium for different values of the gas pressure top boundary condition, $P_{g}(z_{max}=2048\text{ km})$. Left column: Stokes parameters, from top to bottom, I, Q, U, and V, for the simulated case, ${\varmathbb I}^{\text{syn}}(\lambda)$ (solid-black lines), the inversion with $P_{g}^{1}(z_{max}=2048\text{ km})=2.54$ dyn$/\text{cm}^{2}$ (${\varmathbb I}^{\text{inv}}_{1}(\lambda)$, solid red lines) and the inversion with $P_{g}^{1}(z_{max}=2048\text{ km})=254$ dyn$/\text{cm}^{2}$ (${\varmathbb I}^{\text{inv}}_{2}(\lambda)$, dash blue lines). Middle column: Thermodynamic parameters used for the simulated data ($\mathbf{X}^{\text{syn}}(z)$) together with the ones obtained for each inversion ($\mathbf{X}^{\prime}_{1}(z)$ and $\mathbf{X}^{\prime}_{2}(z)$, following the same color code). Right column: Same thermodynamic parameters ($\mathbf{\tilde{X}}^{\text{syn}}(z)$, $\mathbf{\tilde{X}}^{\prime}_{1}(z)$, and $\mathbf{\tilde{X}}^{\prime}_{2}(z)$), but setting a common height origin for the three cases $h(\tau_{5}=1)=1000$ km. In the upper-center and -right panels we include  horizontal lines with ticks for, from left to right, the heights at which \mbox{$\lg\tau_{5}=0,-1,-2,-3$} for each atmosphere (following the same color code).}
      \label{Fig:HyEAmb}
\end{figure*}

An important drawback when working in the geometrical scale $z$, compared to optical scale $\tau$, is that there are two intrinsic degeneracies involving the thermodynamic properties. 1) Any shift in height of the whole atmosphere leads to exactly the same Stokes spectra. 2) There are different combinations of the thermodynamic parameters that can lead to the same absorption coefficient. For instance, one could increase the density (increasing the number of atoms) and at the same time decrease the temperature (decreasing the population of an ionization stage) so that the absorption coefficient remains the same. To better illustrate this second point (as the first one is straightforward), Fig. \ref{Fig:ProfSynAmb} shows the inversion of a simulated Stokes profile of the same spectral line configuration employed until now, ${\varmathbb I}^{\text{syn}}(\lambda)$, which is obtained through a given atmospheric model (black lines). The strategy is to invert the gas pressure $P_{g}(z)$ while forcing the temperature stratification $T(z)$ to be different (see blue and red color lines) from the one used for the synthesis. This way, if we can fit, up to a given threshold, the Stokes profiles for the different fixed temperature stratifications, it means that there exists more than one combination of thermodynamic parameters as a function of the geometrical scale $z$ that yield the same Stokes profiles. As shown, this is exactly the case, as for the different temperatures the gas pressure and densities have changed (the former through the inversion and the latter through the EOS) so that the Stokes profiles (${\varmathbb I}^{\text{inv}}_{1}(\lambda)$ and ${\varmathbb I}^{\text{inv}}_{2}(\lambda)$) are approximately equally well fit. It is worth noticing that, even for the extreme temperatures provided, the fit of the Stokes profiles is quite good, in particular for the first case (red). The second one (blue) is slightly worse because, as the supplied temperature gradient is steeper than the reference one, the spectral line formation heights are comprised in a smaller number of slabs and so the solution of the RTE is not as accurate as for the simulated case, giving rise to small discrepancies in the stokes profiles, mostly close to the core of the line in any Stokes parameter. The number of free parameters in this test is the same as in the previous one: 256.

 One possibility to circumvent this degeneracy is to assume hydrostatic equilibrium. Here, the momentum equation \citep[see, e.g.,][]{priest1987}, simplifies to $dP_{g}/dz=-\rho g$, where $g$ refers to the surface gravity of the Sun. Under such an assumption and together with the equation of state, the gas pressure and density, $P_{g}(z)$ and $\rho(z)$, can be obtained provided the temperature $T(z)$ plus a boundary condition for the gas pressure. This can be done numerically by means of any one-dimensional solver for first-order differential equations (e.g., Runge-Kutta). By including this additional constraint, one imposes a tight restriction on the thermodynamic stratifications so that only $T(z)$ and a boundary condition for the gas pressure remain as free parameters during the inversion. An equivalent problem can be formulated using the density $\rho(z)$ as free parameter instead of the temperature. This assumption considers the most simplified version of the MHD equation of motion and is often made for RTE solver codes working in optical depth $\tau$ such as SIR \citep{ruizcobo1992}, SPINOR \citep{frutiger2000}, and NICOLE \citep{socasnavarro2015}. 

In Fig. \ref{Fig:HyEAmb} we test the uniqueness of the solution in the thermodynamic parameters using the additional restriction of hydrostatic equilibrium described above. To do so we supplied a temperature stratification $T(z)$ as well as a value for the gas pressure in the uppermost atmospheric layer $P_{g}(z_{max})=25.4$ dyn/cm$^2$, with $z_{max}=2048$ km, and calculate the gas pressure and density stratifications, $P_{g}(z)$ and $\rho(z)$ using hydrostatic equilibrium. The resulting atmosphere, which we use to obtain simulated Stokes profiles, ${\varmathbb I}^{\text{syn}}(\lambda)$, are denoted by black lines in Fig. \ref{Fig:HyEAmb}. Once this atmosphere and its resulting Stokes profiles are set, we perform different inversions for the temperature stratifications, $T(z)$, under the assumption of hydrostatic equilibrium, while fixing different values for the gas pressure at the top boundary, $P^{1}_{g}(z_{max})=2.54$ dyn/cm$^2$ and $P^{2}_{g}(z_{max})=254$ dyn/cm$^2$. Since only one physical parameter is inverted for all available heights, the number of free parameters is still 256. If ${\varmathbb I}^{\text{syn}}(\lambda)$ is successfully fitted, this readily implies that the temperature $T(z)$ is able to compensate for the different values of the boundary condition in the gas pressure. As is seen in the middle row of Fig. \ref{Fig:HyEAmb}, this is exactly the case, as there are several thermodynamic stratifications that fit the spectra equally well, up to an acceptable error.

\begin{figure*}
\centering
\includegraphics[width=\hsize]{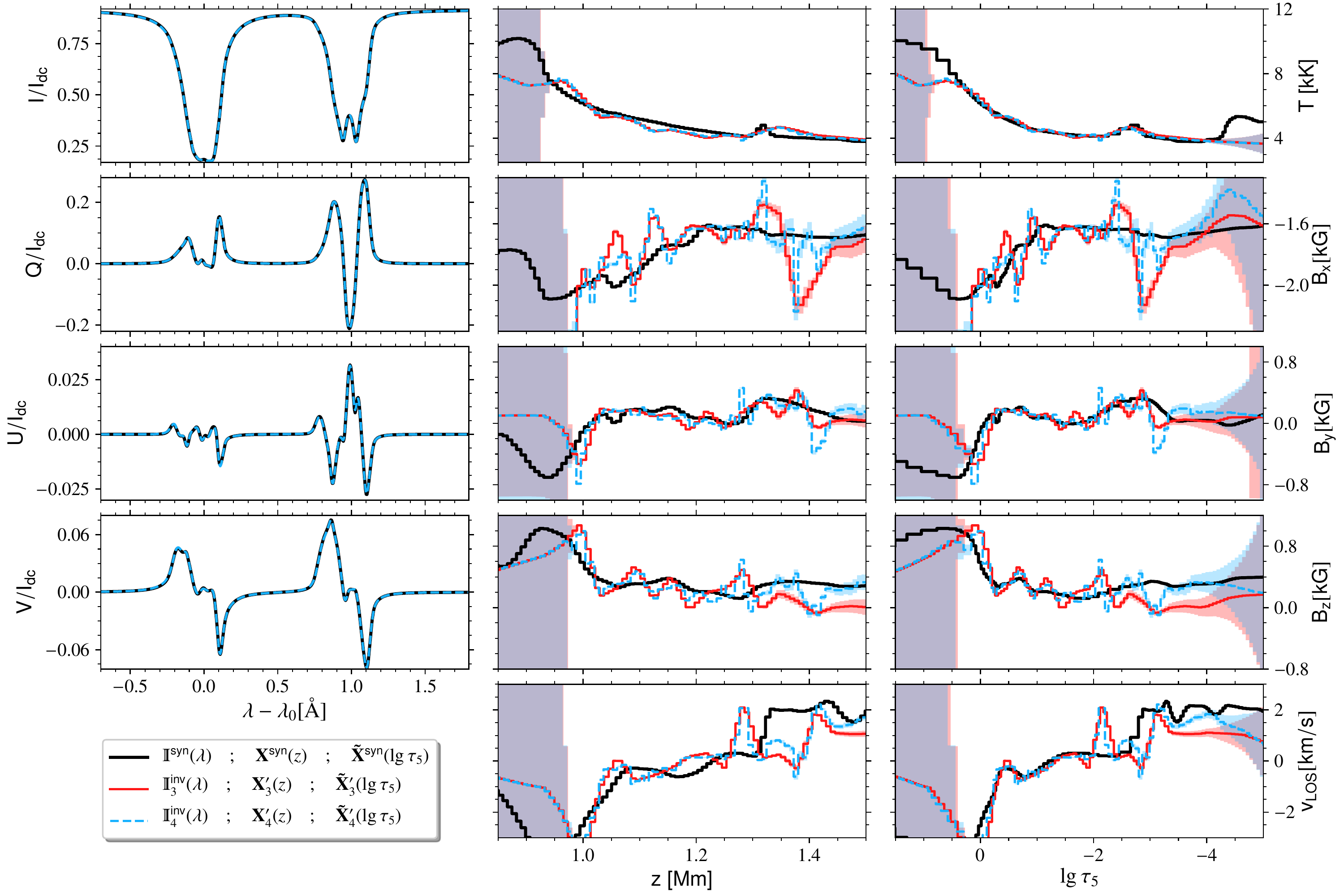}
   \caption{Inversion of a penumbral pixel (highlighted in Fig. \ref{Fig:ProfSyn} with a red square). In black, the simulated Stokes parameters, ${\varmathbb I}^{\text{syn}}(\lambda)$, (left column) and atmosphere model parameters inverted in geometrical scale (${\mathbf X}^{\text{syn}}(z)$, middle column) and in optical depth scale (${\mathbf X}^{\text{syn}}(\lg\tau_{5})$, rightmost column). The results of the inversion in hydrostatic equilibrium for the third and fourth cycle (see text), ${\varmathbb I}^{\text{inv}}_{3}(\lambda)$) and ${\varmathbb I}^{\text{inv}}_{4}(\lambda)$), are in solid red and dashed blue, respectively. The physical parameters inverted are $T(z)$, $v_{LOS}(z)$, $B_{x}(z)$, $B_{y}(z)$, and $B_{z}(z)$. Since the height origin is not absolute (see Sect. \ref{sec:inverse_solver_results}), each atmosphere model is forced to have $h(\tau_{5}=1)=1000 $ km. The atmosphere models are only shown between $lg(\tau_{5})=1.5$ and $lg(\tau_{5})=-5$ as the Stokes profiles, given the spectral line setup, are sensitive inside these optical depths, though the whole atmosphere is inverted. The shadowed area shows the lower limit errors for each physical parameter.}
      \label{Fig:InvWhl}
\end{figure*}

We note that, in spite of the seemingly very different stratifications of the thermodynamic parameters (middle panels in Fig. \ref{Fig:HyEAmb}), they agree well if shifted to share a common height origin. In Fig. \ref{Fig:HyEAmb} (rightmost panels), we perform a shift of the $z$-scale such that the geometrical height at which the $\tau_{5}=1$ forms is the same in all three (black --$\mathbf{\tilde{X}}^{\text{syn}}(z)$-- and blue and red --$\mathbf{\tilde{X}}^{i}(z)$--) curves. The shifts applied in this case are $\Delta z^{1}=83$ and $\Delta z^{2}=-228$ km, for the cases of $P_{g}^{1}$ and $P_{g}^{2}$, respectively, as the top boundary conditions in the gas pressure. This result implies that the assumption of hydrostatic equilibrium breaks the aforementioned degeneracy regarding the thermodynamic parameters (point {\it 2} in Sect. \ref{sec:inverse_solver_results}), and allows us to infer the correct temperature stratification $T(z)$ (as long as $P_{g}$ and $\rho$ are obtained in hydrostatic equilibrium as well), save for a shift in the absolute reference for the origin of the $z$-scale. This was expected since, as we explained above (point {\it 1}), any shift in height of the atmosphere model as a whole leads to exactly the same Stokes parameters. Setting a common origin would require taking into account additional terms in the momentum equation, in addition to the hydrostatic one, such as the magnetic pressure and tension terms \citep[see Sect. 2.7 of][]{priest1987}. This cannot be done in a one-dimensional approach like the one used in this paper, where each pixel ($x$,$y$) on the solar surface is treated independently, and therefore setting a common origin lies out of the scope of the present investigation. Yet, some general scenarios are discussed later.

Finally, we perform a similar inversion as the one we have just carried out but this time applied to the Stokes profiles coming from the same snapshot shown in Fig. \ref{Fig:ProfSyn}. Here the physical parameters used to produce the simulated Stokes pseudovector, ${\varmathbb I}^{\text{syn}}(\lambda)$, do not rely on the assumption of hydrostatic equilibrium, but rather they verify the full set of MHD equations. Therefore, the inversion, assuming that hydrostatic equilibrium applies, will provide insight into how accurate the inferences of the various height-dependent parameters are. Contrary to all previous tests, we now include in the inversion the three components of the magnetic field, $B_{x,y,z}$, as well as the LOS velocity $v_{LOS}$ and we do not force them to be constant in height. The inversion is performed for the whole height scale $z$ grid and thus there are 1280 free parameters.

We consider the same pixel as before (red square in Fig. \ref{Fig:ProfSyn}). The physical parameters (${\mathbf X}^{\text{syn}}(z)$) together with the Stokes simulated spectra they give to (${\varmathbb I}^{\text{syn}}(\lambda)$) are shown in Fig. \ref{Fig:InvWhl} (black lines). The results of the inversion for the temperature $T(z)$, the LOS velocity $v_{LOS}$, and the three cartesian components of the magnetic field are shown in red (${\mathbf X}^{\prime}(z)$). We recall here that we employ two different approaches to deal with the large number of free parameters involved in the inverse problem (1280). On the one hand, the inversion is performed in a sequential manner. Thus, in a first step or cycle, each physical quantity inverted ($T$, $v_{LOS}$, and $B_{x,y,z}$) was modified by a constant value across all $z$-slabs. In the following steps or cycles, the perturbation for each parameter was linearly interpolated from the calculated ones at 8, 64, and finally all 256 different heights (equally spaced in z). On the other hand, the regularization introduced during the inversion of the dampened Hessian, (highly) reduces the dimension of the problem. For instance, in the last step of the inversion, there are 256 free parameters per physical quantity, but as is evident from middle column in Fig. \ref{Fig:InvWhl}, perturbations in slabs below 900 km are ignored by the SVD method as the eigenvalues of the dampened Hessian associated to these heights are below the $10^{-4}$ threshold. Something similar applies to higher layers where the density is so small that Stokes profiles are highly insensitive to the physical parameters on these slabs. In addition to these, the regularization also implies a smoothing of the solution in the sense that it reduces the effective height resolution. The smaller the threshold the larger the smoothing. For the inversion shown in Fig. \ref{Fig:InvWhl}, of the 1280 possible free parameters only about 350 are really considered in the inversion after applying a SVD threshold of $10^{-4}$. The result, as can be seen from the agreement between the simulated (black lines on the leftmost column in Fig. \ref{Fig:InvWhl}) and fitted (red lines in the same panels) Stokes profiles, is good.

\begin{figure}
\centering
\includegraphics[width=0.98\hsize]{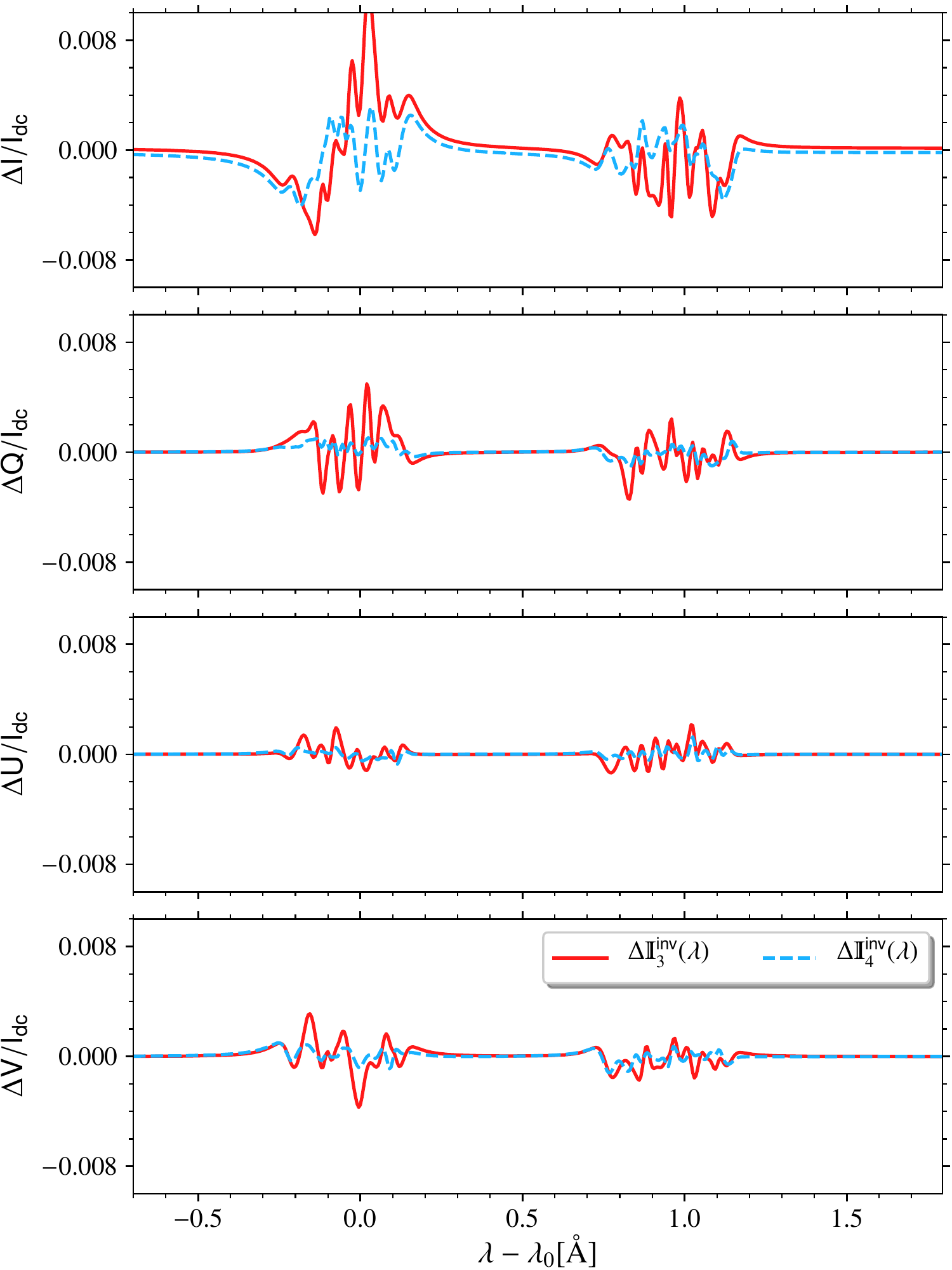}
   \caption{Difference between the Stokes profiles of the inversion results and the inverted ones for the example considered in Fig. \ref{Fig:InvWhl} (following the same color code).}
      \label{Fig:InvWhldiff}
\end{figure}

If we focus on the physical parameters we first see that inferred parameters sometimes show large variations between slabs, mostly in LOS velocity and the magnetic field components. The reason is that since we are using the original height grid from the MHD simulation, the effect of a single slab over the overall Stokes profile appearance might be tiny, and so large excursions are possible with this inversion setup. Also, we see that the assumption of hydrostatic equilibrium leads to a general misfit of the $z$ {\textit{scale}} (middle row in Fig. \ref{Fig:InvWhl}), that is mostly seen as a different slope between the simulated and inferred temperature and as a compression of the inferred (red) LOS velocity and magnetic field components with respect to the simulated ones (black). We notice that, as mentioned before, there is no absolute height-scale-origin and therefore we set the height at which $\tau_{5}=1$  as 1000 km. In contrast, the rightmost column of Fig. \ref{Fig:InvWhl} shows that, in optical depth scale and in between $\lg\tau_{5}=0$ and $\lg\tau_{5}=-3$, the physical parameters are nicely inferred. It is interesting to point out that the excellent recovery of the optical depth stratification is the result of the combined action of the two points mentioned above. First, if there is no regularization applied while calculating the inverse of the dampened Hessian, there is no convergence at all as the inverted matrix is (almost certainly for this large number of free parameters) quasi-singular.  Second, without the sequential approach to the solution by progressive increase of the number of free parameters, the inversion results are strongly determined by the initial guess model. In this sense, the first inversion steps or cycles with a reduced number of free parameters can be seen as a way to get a good enough guess for the final one, as they do not capture all the complexity of the profiles but they are better suited to retrieve the general atmospheric behavior. For instance, in Fig. \ref{Fig:InvWhl}, we have included in red and blue the final atmospheric model for the third and fourth cycles of the inversion, respectively (i.e., the ones with 64 and 256 perturbations per physical parameter). The results after the third cycle are already very close to the real one (black lines) but there are still some features that are slightly improved in the fourth step or `cycle' of the inversion, where the perturbation of the model atmosphere is calculated over the whole geometrical scale grid. The sequential increase of the number of free parameters must be carried out with caution when applying the inversion to real observations. This is illustrated in Fig. \ref{Fig:InvWhldiff}, where the differences between the fitted and simulated profiles after the third and fourth inversion cycles are displayed (following the same color code as in Fig. \ref{Fig:InvWhl}). These differences are already close to $\sim10^{-3},10^{-4}I_{c}$  (typical noise level in modern spectropolarimeters) after the third cycle (solid red lines) and therefore a small number of free parameters might be kept in real applications.
 
One may wonder as to why the model atmosphere in optical depth is so good while in geometrical height it is not. The reason is that in order to fit the Stokes profiles (${\varmathbb I}^{\text{syn}}(\lambda)$), the temperature, magnetic field components and LOS velocity in optical depth must be the same (up to a given accuracy) to that used for the synthesis. Yet, since hydrostatic equilibrium involves a very strong restriction, as it ignores all the other forces present in the MHD equation of motion, the gas pressure and density obtained in this fashion are unrealistic and so is the continuum opacity that relates geometrical and optical depth scales. Thus, in geometrical scale, the various physical parameters inverted, are inaccurately inferred. Here, it is clear that the ability to determine the various physical parameters in a three-dimensional volume from a set of polarimetric observations relies on the accuracy with which the gas pressure is known or estimated.
 
\section{Conclusions and discussion}

We present a numerical code that solves the RTE for polarized light (in the Zeeman regime) in geometrical scale $z$ in a one-dimensional setup, which is made publicly available\footnote{\url{https://github.com/apy-github/firtez-dz}}. It includes, for the first time, the analytical calculation of the Response functions in height scale $z$, which allows a better understanding as to how the optical depth, and as a consequence the spectral lines, are formed in $z$. It is also possible to solve the inverse problem, inferring all relevant physical parameters from a given observation. The code is parallelized so that large datasets can be easily tackled both in the forward and the inverse solver. Here, we have tested the performance of the inversion using a huge number of free parameters, yet this number can be easily reduced if desired.

We see that solving the RTE in geometrical scale $z$ yields, when facing the inverse problem, to two intrinsic degeneracies: 1) the absence, when analyzing each pixel independently, of an absolute height scale origin, and 2) solving the RTE in geometrical scale leads to a degeneracy concerning the thermodynamical parameters: $T(z)$, $P_{g}(z)$, and $\rho(z)$. The latter appears because there are different combinations of these physical parameters that yield the same absorption coefficient. In other words, there is no guarantee that, provided an observation, different attempts at the simultaneous inversion of two thermodynamical parameters will lead to the same geometrical scale $z$ stratification.

One possibility to avoid this multiplicity on the solution of the inverse problem is to include an additional constraint such as hydrostatic equilibrium. Under this assumption, provided a boundary condition for the gas pressure, $P_{g}(z_{max})$, and either $T(z)$ or $\rho(z)$, the remaining stratification of the thermodynamic parameters can all be obtained without degeneracy. The inclusion of this constraint in the solution of the inverse problem guarantees the uniqueness of inferred stratifications as a function of z, except for an absolute height-scale origin. The absence of such an origin cannot be addressed in a one-dimensional approach, as it requires that additional terms be considered (e.g., magnetic pressure and tension), in addition to the hydrostatic one, in the momentum equation. This is evident as the Lorentz force involves the calculation of spatial derivatives that can only be evaluated in a three-dimensional approach.

Finally, we tested the effects of the assumption of hydrostatic equilibrium on the inference of height-dependent physical parameters $\mathbf{X}(z)$. We see that since the gas pressure obtained under this assumption is not very accurate, the inferred $z$-dependences of the physical parameters deviate from the original ones. However, the inference of the temperature, the three components of the magnetic field vector, and LOS velocity is accurate in the optical depth scale $\tau$.

These two open problems, namely the lack of an absolute height origin when treating each pixel individually (i.e., one-dimensional approach), and the unreliability of the inferred gas pressure and density (in both the $z$ and $\tau$ scales), can be addressed by including more terms, in addition to the hydrostatic one, in the momentum equation. This would allow a more realistic gas pressure, $P_{g}(z)$, and density $\rho(z)$ to be retrieved. Once this is achieved, as we demonstrate here, all other physical parameters (including the magnetic field) can be reliably determined as a function of the geometrical height $z$, which in turn would allow a more realistic determination of electric currents, $\mathbf{j}=\frac{1}{4\,\pi}\nabla\times\mathbf{B}$, and to infer a proper solenoidal magnetic field $\nabla\cdot\mathbf{B}=0$. Since the proper determination of the currents and magnetic fields relies on the accuracy of the $P_{g}(z)$ and, at the same time, the latter depends on the currents and magnetic fields, this process will require an iterative approach. This would represent a major leap in the study of the magnetic topology in the solar atmosphere.

In the past, some attempts have been made with this aim \citep{solanki1993,pillet1993,mathew2004}, but it remains unclear how to proceed from observations to three-dimensional physical parameters. An ambitious possibility, presented by \cite{riethmueller2017}, is to directly use state-of-the-art MHD codes to iteratively retrieve a simulation run that gets closer and closer to an observation. This way, since one is already solving the whole three-dimensional physical volume consistently, the final atmosphere has taken into account all the information available. However, this approach is computationally expensive, as it requires running MHD simulations and there is no guarantee that the simulation reproduces all the observed features. A different approach was presented by \cite{puschmann2010} where they inverted an area of the solar surface under a one-dimensional approach and then included additional constraints by means of the hydrodynamic and magnetic terms of the MHD equation together with the $\nabla\cdot\mathbf{B}=0$ condition. However, in this approach the new and final three-dimensional atmosphere is not fully compatible with the observations. In addition, since it relies on evaluations of vertical displacements in each pixel, for large fields-of-views the number of free parameters becomes extremely large. A possibility to circumvent the latter was presented by \cite{loptien2018}, where they used a modification of the method of \cite{puschmann2010}, only accounting for the $\nabla\cdot\mathbf{B}=0$ condition and using a Fourier decomposition so that the free parameters can be taken to a workable size. In this work, we present a numerical code that allows this problem to be faced in a different way, yet follows the idea presented by \cite{puschmann2010}, namely including the hydrodynamic and magnetic terms of the MHD equation and the $\nabla\cdot\mathbf{B}=0$ condition so that the three-dimensional atmosphere is  horizontally consistent. Here however there are two advantages: 1) the horizontal derivatives required that the additional terms included in the equation of motion can be directly handled, and 2) the results of solving these additional equations, which lead to a different gas pressure, can be used to feed again the inversion code to retrieve a three-dimensional atmosphere that would be fully consistent with the observations.


\begin{acknowledgements}
We thank the referee for many interesting suggestions that improved the paper. This work was supported by the German \emph{Deut\-sche For\-schungs\-ge\-mein\-schaft, DFG\/} project number 321818926. JMB acknowledges financial support from the Spanish Ministry of Economy and Competitiveness (MINECO) under the 2015 Severo Ochoa Program MINECO SEV-2015-0548.

\end{acknowledgements}

%
   \bibliographystyle{aa} 
   \bibliography{biblio} 
%

\begin{appendix}
\section{Derivative of the number of electron density with respect to temperature at constant gas density}
\label{App:dnedt_rho}

We focus on the derivative of the density of electrons with respect to the temperature at constant gas pressure. The number of electron density is given by the sum over the whole set of atomic elements (index $i$) and for the ionization levels (index $j$):

\begin{equation*}
  n_{e}=\sum_{i}\sum_{j=1}^{N_{ion}}C_{i,j}\,n_{i,j},
\end{equation*}

where $n_{i,j}$ gives the number of atoms per cm$^{3}$ of the species $i$ at the ionization stage $j$. In the current implementation of the code, $N_{ion}=3$ and $C_{i,j}$ encodes the additional electrons freed (absorbed) by each species at each ionization stage. For hydrogen, $j=1$ refers to the negative hydrogen ion so $C_{1,1}=-1$, $j=2$ is the neutral hydrogen ($C_{1,2}=0$), and $j=3$ is the ionized hydrogen ($C_{1,3}=1$). For the rest of elements, $j=1$ is the neutral atom ($C_{i\ne 1,1}=0$) and $j=2$ and $j=3$ refer to the once and twice ionized atoms ($C_{i\ne 1,2}=C_{i\ne 1,3}=1$). Therefore, the derivative remains as:

\begin{equation}
  \left(\frac{\partial n_{e}}{\partial T}\right)_{\rho}=
  \left(\frac{\partial }{\partial T}\right)_{\rho}\left[\sum_{i}\sum_{j=1}^{3}C_{i,j}\,n_{i,j}\right].
  \label{eq:toderiv}
\end{equation}
 
Now $n_{i,j}$ is given by the product of the density of $i$ atoms multiplied by the fraction of these atoms in the ionization state $j$:

\begin{equation}
  n_{i,j}=n_{1}\,10^{A_{i}-12}f_{i,j},
  \label{eq:toderiv_aux1}
\end{equation}

where $n_{1}$ is the density of hydrogen atoms and $A_{i}$ the abundance of species $i$, which is defined in the usual astronomical scale: $A_{i}=lg(n_{i}/n_{H})+12$. We now refer to $f_{i,j}$ as the fraction of atoms of the species $i$ at the ionization state $j$ with respect to the total number of the same species i:

\begin{equation}
  f_{i,j}=\frac{m_{i,j}}{\sum_{k=1}^{3}m_{i,k}},
  \label{eq:toderiv_aux2}
\end{equation}

where we assume that there is no element that is ionized more than three times. Since we are assuming LTE, the various ionization states are given by the Saha equation, so it can be seen that the population of each ionization state is:

\begin{equation}
\begin{split}
m_{i,1}&=n_{e}^{2}\alpha_{i,1}\alpha_{i,2},\\
m_{i,2}&=n_{e}\alpha_{i,2},\\
m_{i,3}&=1,
\end{split}
\label{eq:toderiv2_aux2}
\end{equation}

where

\begin{equation}
\alpha_{i,j}=\frac{u_{i,j}}{2\,u_{i,j+1}}\,\frac{c_{1}}{T^{3/2}}\,\exp{\frac{\epsilon_{i,j}}{k_{B}\,T}}.
\label{eq:alphaterms}
\end{equation}

Here, $u_{i,j}$ refers to the partition function of the species $i$ at the ionization state $j$, \mbox{$c_{1}=\frac{h^{3}}{(2\,\pi\,m_{e}\,k_{B})^{3/2}}$}, $\epsilon_{i,j}$ is the ionization potential, and $h$, $m_{e}$, and $k_{B}$ are the Planck, the electron mass, and the Boltzmann constants, respectively.

 Now, one can substitute Eqs. \ref{eq:toderiv_aux1} and \ref{eq:toderiv_aux2} into Eq. \ref{eq:toderiv} and get

\begin{equation}
\begin{split}
  \left(\frac{\partial n_{e}}{\partial T}\right)_{\rho}=
  \sum_{i}\sum_{j=1}^{3}&C_{i,j}\,10^{A_{i}-12}\Bigg\{
  \left(\frac{\partial n_{1}}{\partial T}\right)_{\rho}f_{i,j}\\
  &+n_{1}\left(\frac{\partial }{\partial T}\right)_{\rho}\Bigg[
  \frac{m_{i,j}}{\sum_{k=1}^{3}m_{i,k}}\Bigg]
  \Bigg\},
\end{split}
\label{eq:toderiv2}
\end{equation}

where we can substitute $n_{1}$, provided that our equation of state is the ideal gas equation, so that its derivative is

\begin{equation}
  \left(\frac{\partial n_{1}}{\partial T}\right)_{\rho}=
  \left(\frac{\partial }{\partial T}\right)_{\rho}
  \left[\frac{1}{s^{\prime}}\left(\rho-n_{e}\,m_{e}\right)\right]
  =\frac{-m_{e}}{s^{\prime}}\left(\frac{\partial n_{e}}{\partial T}\right)_{\rho},
\label{eq:toderiv2_aux1}
\end{equation}

where \mbox{$s^{\prime}=\sum_{i}10^{A_{i}-12}m_{i}$}. We focus now on the last derivative appearing in Eq. \ref{eq:toderiv2} and we define $d_{i}=\sum_{k=1}^{3}m_{i,k}$ so that

\begin{equation}
\begin{split}
\left(\frac{\partial }{\partial T}\right)_{\rho}\Bigg[
  \frac{m_{i,j}}{\sum_{k=1}^{3}m_{i,k}}\Bigg]&=
  \left(\frac{\partial }{\partial T}\right)_{\rho}\Bigg[
  \frac{m_{i,j}}{d_{i}}\Bigg]=\\
&=\left(\frac{\partial m_{i,j}}{\partial T}\right)_{\rho}\,d_{i}^{-1}-m_{i,j}\,\left(\frac{\partial d_{i}}{\partial T}\right)_{\rho}\,d_{i}^{-2}.
\end{split}
\label{eq:toderiv2_aux3}
\end{equation}

From Eq. \ref{eq:toderiv2_aux2}, it is evident that both derivatives on the right-hand side of the Eq. \ref{eq:toderiv2_aux3} depend on the derivative we want to calculate ($\left(\frac{\partial n_{e}}{\partial T}\right)_{\rho}$). Let us write $\left(\frac{\partial m_{i,j}}{\partial T}\right)_{\rho}$ and $\left(\frac{\partial d_{i}}{\partial T}\right)_{\rho}$ as a sum of a part that depends on $\left(\frac{\partial n_{e}}{\partial T}\right)_{\rho}$ and another that does not depend on it:

\begin{equation}
  \begin{split}
  \left(\frac{\partial d_{i}}{\partial T}\right)_{\rho}=&\mathcal{D}_{i}
  +\mathcal{F}_{i}\left(\frac{\partial n_{e}}{\partial T}\right)_{\rho}\\
  \left(\frac{\partial m_{i,j}}{\partial T}\right)_{\rho}=&
  \mathcal{G}_{i,j}+\mathcal{H}_{i,j}\left(\frac{\partial n_{e}}{\partial T}\right)_{\rho},
  \end{split}
  \label{eq:toderiv2_aux4}
\end{equation}

where the definitions for $\mathcal{D}_{i}$, $\mathcal{F}_{i}$, $\mathcal{G}_{i,j}$, and $\mathcal{H}_{i,j}$ are detailed below. If we introduce Eq. \ref{eq:toderiv2_aux4} into Eq. \ref{eq:toderiv2_aux3} and then this latter together with Eq. \ref{eq:toderiv2_aux1} into \ref{eq:toderiv2}. These, we obtain

\begin{equation}
  \begin{split}
  \left(\frac{\partial n_{e}}{\partial T}\right)_{\rho}=&
  \frac{-m_{e}\,n_{e}}{s^{\prime}\,n_{1}}
  \left(\frac{\partial n_{e}}{\partial T}\right)_{\rho}\\
  &+\sum_{i}^{N}n_{i}\sum_{j}^{3}C_{i,j}\Bigg[
  \Bigg(\mathcal{G}_{i,j}+\mathcal{H}_{i,j}
  \left(\frac{\partial n_{e}}{\partial T}\right)_{\rho}\Bigg)\,d_{i}^{-1}\\
  &-m_{i,j}\,\left(\mathcal{D}_{i}+\mathcal{F}_{i}\left(\frac{\partial n_{e}}{\partial T}\right)_{\rho}\right)\,d_{i}^{-2}\Bigg].
  \end{split}
\end{equation}

We can now group terms with $\left(\frac{\partial n_{e}}{\partial T}\right)_{\rho}$ and get the analytical expression for it:

\begin{equation}
  \begin{split}
  \left(\frac{\partial n_{e}}{\partial T}\right)_{\rho}&
  =\frac{\sum\limits_{i}^{N}n_{i}\sum\limits_{j}^{3}C_{i,j}\left[\mathcal{G}_{i,j}\,d_{i}^{-1}
  -m_{i,j}\,\mathcal{D}_{i}\,d_{i}^{-2}\right]}
  {1+\frac{m_{e}}{s^{\prime}}\,\frac{n_{e}}{n_{1}}
  \sum\limits_{i}^{N}n_{i}\sum\limits_{j}^{3}C_{i,j}\left[\mathcal{H}_{i,j}\,d_{i}^{-1}
  -m_{i,j}\,\mathcal{F}_{i}\,d_{i}^{-2}\right]}\\
  \end{split}
.\end{equation}

This allows the calculation of the derivative of the number of electron density with respect to the temperature at constant density. We just need to specify the expressions for: $\mathcal{D}_{i}$, $\mathcal{F}_{i}$, $\mathcal{G}_{i,j}$, and $\mathcal{H}_{i,j}$. If we write

\begin{equation}
  m_{i,j}=\mathcal{A}_{i,j}\,n_{e}^2+\mathcal{B}_{i,j}\,n_{e}+\mathcal{C}_{i,j},
\end{equation}

then

\begin{equation}
  \begin{split}
  \mathcal{G}_{i,j}=&\left(\frac{\partial \mathcal{A}_{i,j}}{\partial T}\right)_{\rho}\,n_{e}^{2}+\left(\frac{\partial \mathcal{B}_{i,j}}{\partial T}\right)_{\rho}\,n_{e}+\left(\frac{\partial \mathcal{C}_{i,j}}{\partial T}\right)_{\rho}\\
  \mathcal{H}_{i,j}=&2\,\mathcal{A}_{i,j}\,n_{e}+ \mathcal{B}_{i,j},
  \end{split}
\end{equation}

where the various terms for different ionization stages are gathered in Table. \ref{Tab:derivativesTterms}, and

\begin{table}
\caption{Actual mathematical expressions for $\mathcal{A}_{i,j}$, $\mathcal{B}_{i,j}$, $\mathcal{C}_{i,j}$, $\mathcal{G}_{i,j}$, and $\mathcal{H}_{i,j}$ terms for the various ionization stages. These expressions refer to the case of the derivative of the number of electron density with respect to temperature at constant gas density. $\delta$ refers to the partial differentiation with respect to the temperature at constant density.}
\begin{center}
\begin{tabular}{|l|c|c|c|}
 & $j=1$ & $j=2$ & $j=3$ \\
\hline
$\mathcal{A}_{i,j}$ & $\alpha_{i,1}\alpha_{i,2}$ & $0$ & $0$ \\
$\mathcal{B}_{i,j}$ & $0$ & $\alpha_{i,2}$ & $0$ \\
$\mathcal{C}_{i,j}$ & $0$ & $0$ & $1$\\
$\mathcal{G}_{i,j}$ & $n_{e}^{2}\left[\delta\alpha_{i,1}\,\alpha_{i,2}+\alpha_{i,1}\,\delta\alpha_{i,2}\right]$& $\delta\alpha_{i,2}\,n_{e}$ & $0$ \\
$\mathcal{H}_{i,j}$ & $2\,\alpha_{i,1}\alpha_{i,2}\,n_{e}$ & $\alpha_{i,2}$ & $0$ \\
\end{tabular}
\end{center}
\label{Tab:derivativesTterms}
\end{table}


\begin{equation}
  \begin{split}
  \mathcal{F}_{i}&=2\,\alpha_{i,1}\alpha_{i,2}\,n_{e}+\alpha_{i,2}\\
  \mathcal{D}_{i}&=\left(\frac{\partial \alpha_{i,1}}{\partial T}\right)_{\rho}\alpha_{i,2}\,n_{e}^2
+\alpha_{i,1}\left(\frac{\partial \alpha_{i,2}}{\partial T}\right)_{\rho}\,n_{e}^2
+\left(\frac{\partial \alpha_{i,2}}{\partial T}\right)_{\rho}\,n_{e}
  \end{split}
.\end{equation}

\section{Derivative of the number of electron density with respect to gas pressure at constant temperature}
\label{App:dnedpg_t}

For this derivation, we can re-use the first equations appearing in Appendix \ref{App:dnedt_rho} up to Eq. \ref{eq:toderiv2} substituting $\left(\frac{\partial }{\partial T}\right)_{\rho}$ with $\left(\frac{\partial }{\partial P_{g}}\right)_{T}$ and so

\begin{equation}
\begin{split}
  \left(\frac{\partial n_{e}}{\partial P_{g}}\right)_{T}=
  \sum_{i}\sum_{j=1}^{3}&C_{i,j}\,10^{A_{i}-12}\Bigg\{
  \left(\frac{\partial n_{1}}{\partial P_{g}}\right)_{T}f_{i,j}\\
  &+n_{1}\left(\frac{\partial }{\partial P_{g}}\right)_{T}\Bigg[
  \frac{m_{i,j}}{\sum_{k=1}^{3}m_{i,k}}\Bigg]
  \Bigg\}.
\end{split}
\label{eq:toderiv2b}
\end{equation}

The derivative can be calculated making use of the equation of state:

\begin{equation}
  \begin{split}
\left(\frac{\partial n_{1}}{\partial P_{g}}\right)_{T}&=\left(\frac{\partial}{\partial P_{g}}\right)_{T}\left[\frac{1}{s}\left(\frac{P_{g}}{k_{B}\,T}-n_{e}\right)\right]\\
&=\frac{1}{s\,k_{B}\,T}-\frac{1}{s}\left(\frac{\partial n_{e}}{\partial P_{g}}\right)_{T},
  \label{eq:toderiv2b_aux2}
  \end{split}
\end{equation}

and for the second derivative of the right-hand side of Eq. \ref{eq:toderiv2b}, we can follow a similar strategy as in Appendix \ref{App:dnedt_rho}:

\begin{equation}
\begin{split}
\left(\frac{\partial }{\partial P_{g}}\right)_{T}\Bigg[
  \frac{m_{i,j}}{\sum_{k=1}^{3}m_{i,k}}\Bigg]&=
  \left(\frac{\partial }{\partial P_{g}}\right)_{T}\Bigg[
  \frac{m_{i,j}}{d_{i}}\Bigg]=\\
&=\left(\frac{\partial m_{i,j}}{\partial P_{g}}\right)_{T}\,d_{i}^{-1}-m_{i,j}\,\left(\frac{\partial d_{i}}{\partial P_{g}}\right)_{T}\,d_{i}^{-2},
\end{split}
\label{eq:toderiv2b_aux3}
\end{equation}

where we now express the derivatives of the right-hand side as function of a term that depends on $\left(\frac{\partial }{\partial P_{g}}\right)_{T}$ plus another one that does not:

\begin{equation}
  \begin{split}
  \left(\frac{\partial d_{i}}{\partial P_{g}}\right)_{T}=&\mathcal{D}_{i}
  +\mathcal{F}_{i}\left(\frac{\partial n_{e}}{\partial P_{g}}\right)_{T}\\
  \left(\frac{\partial m_{i,j}}{\partial P_{g}}\right)_{T}=&
  \mathcal{G}_{i,j}+\mathcal{H}_{i,j}\left(\frac{\partial n_{e}}{\partial P_{g}}\right)_{T}.
  \end{split}
  \label{eq:toderiv2b_aux4}
\end{equation}

Subsequently, substituting Eq. \ref{eq:toderiv2b_aux4} in Eq. \ref{eq:toderiv2b_aux3} and the latter together with Eq. \ref{eq:toderiv2b_aux2} in Eq. \ref{eq:toderiv2b} we get

\begin{equation}
  \begin{split}
  \left(\frac{\partial n_{e}}{\partial P_{g}}\right)_{T}=&
  \frac{1}{s\,k_{B}\,T}-\frac{1}{s}
  \left(\frac{\partial n_{e}}{\partial T}\right)_{\rho}\\
  &+\sum_{i}^{N}n_{i}\sum_{j}^{3}C_{i,j}\Bigg[
  \Bigg(\mathcal{G}_{i,j}+\mathcal{H}_{i,j}
  \left(\frac{\partial n_{e}}{\partial P_{g}}\right)_{T}\Bigg)\,d_{i}^{-1}\\
  &-m_{i,j}\,\Big(\mathcal{D}_{i}+
  \mathcal{F}_{i}\left(\frac{\partial n_{e}}{\partial P_{g}}\right)_{T}\Big)
  \,d_{i}^{-2}\Bigg].
  \end{split}
\end{equation}

Now we gather together the terms with $\left(\frac{\partial n_{e}}{\partial P_{g}}\right)_{T}$ and solve it, obtaining

\begin{equation}
  \begin{split}
\left(\frac{\partial n_{e}}{\partial P_{g}}\right)_{T}&=
\frac{\frac{1}{s\,k_{B}\,T}}{\Bigg\{1+\frac{1}{s}-\sum_{i}^{N}n_{i}\sum_{j}^{3}C_{i,j}\left[\mathcal{H}_{i,j}\,d_{i}^{-1}-m_{i,j}\,\mathcal{F}_{i}\,d_{i}^{-2}\right]\Bigg\}},
  \end{split}
\end{equation}

where it is to be noted that $\mathcal{G}_{i,j}$ and $\mathcal{D}_{i}$ do not appear, as they are null. We can see this point and retrieve the expressions for $\mathcal{H}_{i,j}$ and $\mathcal{F}_{i}$ if we write

\begin{equation}
  m_{i,j}=\mathcal{A}_{i,j}\,n_{e}^2+\mathcal{B}_{i,j}\,n_{e}+\mathcal{C}_{i,j},
\end{equation}

where $\mathcal{A}_{i,j}$, $\mathcal{B}_{i,j}$, and $\mathcal{C}_{i,j}$ are detailed in Table \ref{Tab:derivativesTtermsB}. Then, taking the derivative of $m_{i,j}$ with respect to gas pressure at constant temperature and bearing in mind Eq. \ref{eq:toderiv2b_aux4} we get

\begin{equation}
  \begin{split}
  \mathcal{G}_{i,j}=&\left(\frac{\partial \mathcal{A}_{i,j}}{\partial P_{g}}\right)_{T}\,n_{e}^{2}+\left(\frac{\partial \mathcal{B}_{i,j}}{\partial P_{g}}\right)_{T}\,n_{e}+\left(\frac{\partial \mathcal{C}_{i,j}}{\partial P_{g}}\right)_{T}\\
  \mathcal{H}_{i,j}=&2\,\mathcal{A}_{i,j}\,n_{e}+ \mathcal{B}_{i,j},
  \end{split}
\end{equation}

where the various terms for different ionization stages are listed in Table \ref{Tab:derivativesTtermsB}. The terms $\mathcal{A}_{i,j}$, $\mathcal{B}_{i,j}$, and $\mathcal{C}_{i,j}$ for the various ionization stages ($j$) are either constants or functions of $\alpha_{i,1}$ and/or $\alpha_{i,2}$. These terms, as can be seen in Eq. \ref{eq:alphaterms}, only depend on the temperature, so their derivatives with respect to gas pressure at constant temperature vanish. In a similar way, it can be shown that by taking the derivative of $d_{i}$ with respect to gas pressure at constant temperature, one leads to

\begin{equation}
  \begin{split}
  \mathcal{F}_{i}&=2\,\alpha_{i,1}\alpha_{i,2}\,n_{e}+\alpha_{i,2}\\
  \mathcal{D}_{i}&=\left(\frac{\partial \alpha_{i,1}}{\partial P_{g}}\right)_{T}\alpha_{i,2}\,n_{e}^2
+\alpha_{i,1}\left(\frac{\partial \alpha_{i,2}}{\partial P_{g}}\right)_{T}\,n_{e}^2
+\left(\frac{\partial \alpha_{i,2}}{\partial P_{g}}\right)_{T}\,n_{e}=0,
  \end{split}
\end{equation}

where these derivatives vanish as $\alpha_{i,1}$ and $\alpha_{i,2}$ only depend on the temperature.

\begin{table}
\caption{Actual mathematical expressions for $\mathcal{A}_{i,j}$, $\mathcal{B}_{i,j}$, $\mathcal{C}_{i,j}$, $\mathcal{G}_{i,j}$, and $\mathcal{H}_{i,j}$ terms for the various ionization stages considered. Here the case of the number of electron density derivatives with respect to the gas pressure at constant temperature.}
\begin{center}
\begin{tabular}{|l|c|c|c|}
 & $j=1$ & $j=2$ & $j=3$ \\
\hline
$\mathcal{A}_{i,j}$ & $\alpha_{i,1}\alpha_{i,2}$ & $0$ & $0$ \\
$\mathcal{B}_{i,j}$ & $0$ & $\alpha_{i,2}$ & $0$ \\
$\mathcal{C}_{i,j}$ & $0$ & $0$ & $1$\\
$\mathcal{G}_{i,j}$ & $0$ & $0$ & $0$ \\
$\mathcal{H}_{i,j}$ & $2\,\alpha_{i,1}\alpha_{i,2}\,n_{e}$ & $\alpha_{i,2}$ & $0$ \\
\end{tabular}
\end{center}
\label{Tab:derivativesTtermsB}
\end{table}

\section{Derivative of the number of electron density with respect to density at constant temperature}
\label{App:dnedrho_t}

This derivation is very similar to the previous one (Appendix \ref{App:dnedpg_t}), where one has to use $\left(\frac{\partial }{\partial \rho}\right)_{T}$ instead of $\left(\frac{\partial }{\partial P_{g}}\right)_{T}$ and taking into account that the hydrogen density number derivative is now

\begin{equation}
  \begin{split}
\left(\frac{\partial n_{1}}{\partial \rho}\right)_{T}&=\left(\frac{\partial}{\partial \rho}\right)_{T}\left[\frac{1}{s^{\prime}}\left(\rho-n_{e}m_{e}\right)\right]\\
&=\frac{1}{s^{\prime}}\left[1-m_{e}\left(\frac{\partial n_{e}}{\partial \rho}\right)_{T}\right].
  \label{eq:toderiv3b_aux2}
  \end{split}
\end{equation}

Thus, following the same steps as in Appendix \ref{App:dnedpg_t}, one reaches a similar expression for $\left(\frac{\partial n_{e}}{\partial \rho}\right)_{T}$:

\begin{equation}
  \begin{split}
\left(\frac{\partial n_{e}}{\partial \rho}\right)_{T}&=
\frac{\frac{1}{s^{\prime}}}{\Bigg\{1+m_{e}-\sum_{i}^{N}n_{i}\sum_{j}^{3}C_{i,j}\left[\mathcal{H}_{i,j}\,d_{i}^{-1}-m_{i,j}\,\mathcal{F}_{i}\,d_{i}^{-2}\right]\Bigg\}},
  \end{split}
\end{equation}

where the various terms appearing in the expression can be found in Appendix \ref{App:dnedpg_t}.

\end{appendix}

\end{document}